\documentclass[printer]{aa}

\usepackage{graphicx}
\usepackage[position=top]{subfig}
\usepackage[final]{pdfpages}
\usepackage{amsmath}
\usepackage[utf8]{inputenc}
\usepackage{epsfig,amsmath,amssymb}
\usepackage[varg]{txfonts}

\def\ga{\,\hbox{\hbox{$ > $}\kern -0.8em \lower 1.0ex\hbox{$\sim$}}\,}
\def\la{\,\hbox{\hbox{$ < $}\kern -0.8em \lower 1.0ex\hbox{$\sim$}}\,}
\def\beq{\begin{equation}}
\def\eeq{\end{equation}}

\titlerunning{Magnetic fields of spiral galaxies}
\authorrunning{Ntormousi}

\begin{document}

%
\title{A dynamo amplifies the magnetic field of a Milky-Way-like galaxy}
\author{Evangelia Ntormousi\inst{1,2,3}, Konstantinos Tassis\inst{1,2}, Fabio Del Sordo\inst{1,2}, Francesca Fragkoudi\inst{4} and R\"{u}diger Pakmor\inst{4}}
\date{Received -- / Accepted --}

\institute{
Foundation for Research and Technology (FORTH), 
Nikolaou Plastira 100, Vassilika Vouton
GR - 711 10, Heraklion, Crete, Greece
\\
\and
Department of Physics and ITCP, University of Crete, 71003 Heraklion, Greece
\\
\and
Scuola Normale Superiore, Piazza dei Cavalieri 7, I-56126 Pisa, Italy
\\
\and
Max-Planck Institute for Astrophysics, Karl-Schwarzschild-Straße 1, 85748 Garching, Germany.}

\abstract
{The magnetic fields of spiral galaxies are so strong that they cannot be primordial. Their typical values are
 over one billion times higher than any value predicted for the early Universe. Explaining this immense growth and
 incorporating it in galaxy evolution theories is one of the long-standing challenges in astrophysics.}
{So far, the most successful theory for the sustained growth of the galactic magnetic field is the alpha-omega dynamo. 
This theory predicts a characteristic dipolar or quadrupolar morphology for the galactic magnetic field, which has been observed in external galaxies. However, so far, there has been no direct demonstration of a mean-field dynamo operating in direct, multi-physics simulations of spiral galaxies. We do so in this work. }
{We employ numerical models of isolated, star-forming spiral galaxies that include a magnetized gaseous disk, a dark matter halo, stars, and stellar feedback.
Naturally, the resulting magnetic field has a complex morphology that includes a strong random component. 
Using a smoothing of the magnetic field on small scales, we are able to separate the mean from the turbulent
 component and analyze them individually.}
{We find that a mean-field dynamo naturally occurs as a result of the dynamical evolution of the galaxy and amplifies the magnetic field by an order of magnitude over half a Gyr. Despite the highly dynamical nature of these models, the morphology of the mean component of the field is identical to analytical predictions.}
{This result underlines the importance of the mean-field dynamo in galactic evolution. Moreover, by demonstrating the
 natural growth of the magnetic field in a complex galactic environment, it brings us a step closer to understanding the
 cosmic origin of magnetic fields.}

\keywords{magnetic fields:galaxies}

\maketitle

\section{Introduction}

The origin of galactic magnetic fields is one of the great unanswered questions in astrophysics today. Although primordial magnetic fields cannot exceed $10^{-9}$ G \citep{Rees_1987, Gnedin_2000, Subramanian_2016}, the magnetic fields in spiral galaxies are of the order of a few $\mu$G \citep{Beck_1996}, already at redshift 2 \citep{Bernet_2008}.  This notable growth inevitably points to a process that amplifies the magnetic field inside spiral galaxies.

This amplification process not only shapes the magnetic field, but it can also affect the galaxy's internal dynamics. The observed values of a few $\mu$G indicate that the magnetic energy of a galaxy is in rough equipartition with the turbulent and thermal energies. Therefore, any theory proposed for the magnetic field evolution must take into account the complexity of galaxy evolution and vice versa.

In the context of galaxy evolution, an obvious possibility is the growth of the magnetic field by gravitational collapse. However, this mechanism is probably inefficient because magnetic diffusion can rapidly disperse the amplified field \citep{Brandenburg_2015}. The galactic dynamo theory \citep{Steenbeck_1966, Parker_1971,Stix_1975,Brandenburg_2005} was put forward to alleviate this problem.

The mean-field dynamo \citep{Parker_1971} creates a poloidal magnetic field component from an initially toroidal configuration and vice versa, leading to an overall amplification of the initial field. Imagine an initially toroidal field in a differentially rotating, stratified disk. As magnetic loops rise above the galactic plane, they spin up due to the differential rotation, and eventually reconnect and merge, creating the poloidal component. The source of buoyancy for the magnetic loops can be shocks, turbulence, cosmic ray diffusion, or stellar feedback. In turn, a toroidal magnetic field emerges from the poloidal component as the differential rotation of the disk winds up the magnetic field lines. This process, called the $\alpha$-$\omega$ dynamo, is sustained for as long as the magnetic energy is much lower than the kinetic energy of the galaxy.

There is a plethora of works studying galactic dynamos in simulated environments \citep{Brandenburg_2015}. 
The first direct simulation of dynamo amplification of the magnetic field considered a small patch of the disk \citep{Gressel_2008}. The advantage of this type of setup is that it allows the evolution of the system until the magnetic field saturates at dynamically important values (e.g., \cite{Gent_2013, Bendre_2015}. \citet{Gissinger_2009} demonstrated that the small-scale helicity injection needed to seed a large-scale dynamo was possible even for a full galactic disk, with high enough resolution. More recent work \citep{Rieder_Teyssier_2016,Rieder_Teyssier_2017} showed that turbulence from stellar feedback could amplify the magnetic field in global simulations of dwarf galaxies up to a few percents of the turbulent kinetic energy of the galaxy. Finally, \cite{Steinwandel_2019} proposed a dynamo mechanism to explain the amplification of the magnetic field in global, SPH galaxy simulations that included a circum-galactic medium, based on the resulting power spectra.
However, no numerical model so far has directly demonstrated a mean-field dynamo amplification of the magnetic field in global simulations of galaxy evolution.

Here, for the first time, we have observed an $\alpha$-$\omega$ dynamo in a highly dynamical environment, where all the relevant components of the galaxy (dark matter, gas, magnetic field, and stars) are allowed to evolve self-consistently. The result comes from a set of simulations we call \emph{Amalthea}, aimed at studying the secular co-evolution of galaxies and their magnetic field.

In Section \ref{numerics} of this paper we outline our numerical approach, in Section \ref{sec:results} we present our results, and conclude the paper in Section \ref{sec:discussion}.

\begin{figure}[h!]
   \centering
       \includegraphics[width=\linewidth]{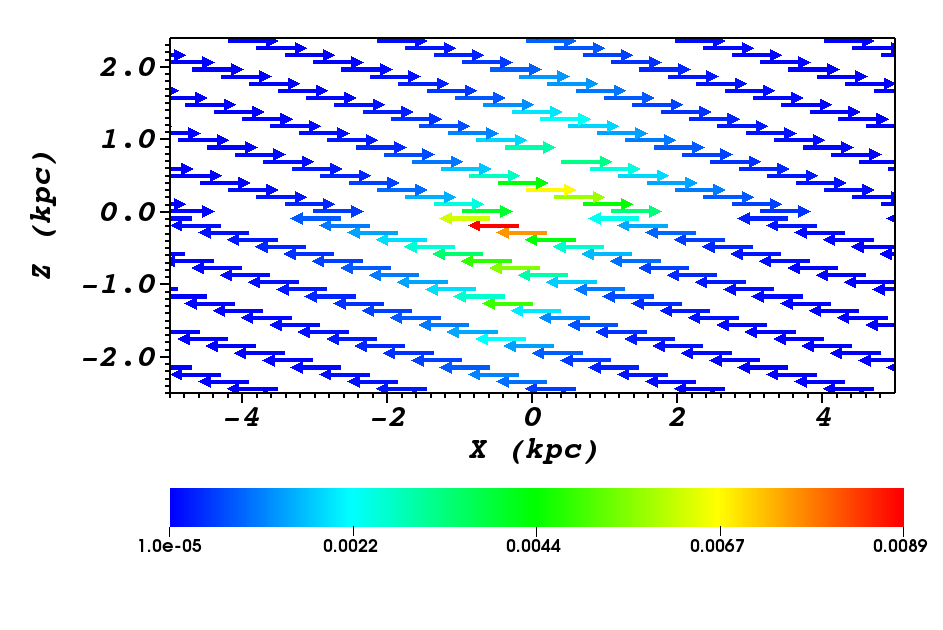}
   \caption{Slice of the initial magnetic field along the xz-plane. The initial field strength drops exponentially with radius and height. The arrows are colored according to the field strength in $\mu$G.}
     \label{initial_conditions}
\end{figure}

\begin{figure*}[h!]
   \centering
       \includegraphics[width=0.7\linewidth]{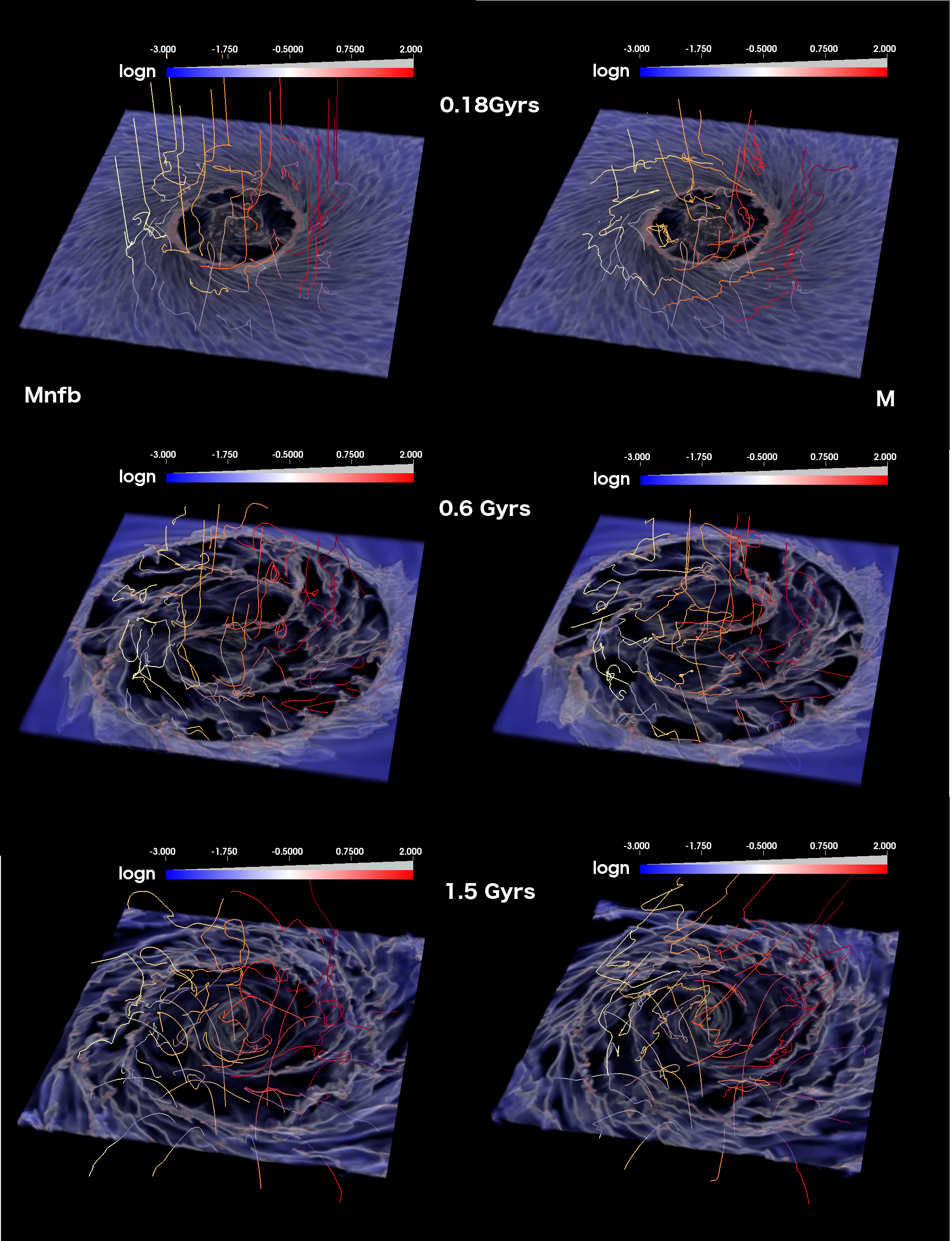}
   \caption{Time evolution of Amalthea Mnfb (left) and Amalthea M (right)  with their magnetic fields. The total gas density is indicated by the colorbars. The magnetic field lines are drawn at the same, regularly spaced locations of the computational volume and are colored according to their starting point. The region shown corresponds to 25 kpc. 
}
     \label{magnetic_evolution}
\end{figure*}
%

\section{Method and setup}
\label{numerics}

Our particular aim with these models is to examine how the magnetic field grows as the galaxy evolves. We simulate isolated, Milky-Way-like
galaxies, using the RAMSES code \citep{Teyssier_02,Fromang_2006}, which allows the simultaneous modeling of the gas and the collisionless components, such as stars and dark matter. This feature is essential for our models.

Table \ref{tab:models} summarizes the properties of the \emph{Amalthea} models \emph{Amalthea M~} is our reference run, and the rest are used to validate our results: \emph{Amalthea Mnfb~} is identical to \emph{Amalthea M~}, but it does not include supernova feedback. \emph{Amalthea Mlr~} is a lower-resolution version of \emph{Amalthea M~}, including one level of refinement less, and \emph{Amalthea W~} has an initial magnetic field ten times smaller than that of the other models.

\subsection{Numerical code}

The simulations were performed with the publicly available code RAMSES .
The MHD equations solved by the code are:

\begin{eqnarray}
\frac{\partial\rho}{\partial t} +\nabla(\rho\bf v) = 0 \label{continuity} \\
\frac{\partial\rho\bf{v}}{\partial t} + \nabla\cdot\left(\rho\bf{v}\bf{v}-\bf{B}\bf{B}\right) + \nabla P_{tot} = -\rho\nabla\phi \label{momentum} \\
\frac{\partial E_{tot}}{\partial t} +\nabla\left[~(E_{tot}+P_{tot})\bf{v} -(\bf{v}\cdot\bf{B})\cdot\bf{B}\right] = -\bf{v}\cdot\nabla\phi -\rho\Lambda + \Gamma \label{energy} \\
\frac{\partial\bf{B}}{\partial t} -\nabla\times(\bf{v}\times\bf{B}) = 0 \label{induction} \\
\nabla\cdot\bf{B} = 0 
\end{eqnarray}
where $\rho$ the gas density, ${\bf{v}}$ the velocity, ${\bf{B}}$ the magnetic field, and $\phi$ the gravitational potential. The cooling and heating rates of the gas as a function of its density and temperature, $\Lambda=\Lambda(\rho,T)$ and $\Gamma=\Gamma(\rho,T)$ are  are calculated according to \citet{Sutherland_Dopita_1993}. $P_{tot}$ is the total pressure:
\begin{equation}
    P_{tot} = P + \frac{\bf{B}\cdot\bf{B}}{2}
\end{equation}
and $E$ the total energy of the fluid:
\begin{equation}
    E = \epsilon + \rho\frac{\bf{v}\cdot\bf{v}}{2} +  + \frac{\bf{B}\cdot\bf{B}}{2}
\end{equation}
where $\epsilon$ is the internal energy of the fluid.
The equation of state for the gas is that of a perfect fluid: $P=(\gamma-1)\epsilon$.

RAMSES uses a constrained transport scheme to evolve the magnetic field, which guarantees $\nabla\cdot{\bf{B}}=0$ always.
The MHD equations are coupled to the stars and dark matter through the gravitational potential (Eqs. \ref{momentum} and \ref{energy}).

\subsection{Setup and initial conditions}

The initial conditions for~the~\emph{Amalthea} models are created using the MCMC-based DICE code \citep{Perret_2014,Perret_2016}.

We use a configuration of stellar and dark matter particles, as well as a hydrodynamical fluid, to simulate a Milky-Way-like galaxy (total
mass~\(M_{tot}=2\cdot10^{12}\ M_{\odot}\)~)~ at redshift z=0. In DICE, we set the virial velocity of the galaxy to 200 km/sec. The mass fraction in dark matter is 98.5\%, represented by 2 million particles, in stars 1.425\%, represented by 1 million particles, and the mass fraction in gas is
0.075\%, represented by 4 million particles.

The dark matter halo follows an NFW profile \citep{NFW96}, with a scale length of 3 kpc and a cutoff of 100 kpc. The stars and gas are
initially placed in exponential disks, with a scale length of 3 kpc and a cutoff of 12 kpc for the stars, and a scale length of 4kpc with a cutoff of 15 kpc for the gas. The gas temperature is set to 10,000 K, with no initial turbulent field.

The initial magnetic field is toroidal, with a scale height and scale length of 1 kpc. The central strength of the field is a free parameter of the magnetic field model. It is set to 0.1 $\mu$G for all the models in this work apart from Model W that starts with a central magnetic field strength of 0.01 $\mu$G. Fig. \ref{initial_conditions} shows a xz-slice of the initial magnetic field configuration, illustrating the symmetry of the initial conditions.

Star formation is simulated by forming sink particles when the density exceeds 1 cm$^{-3}$ and the local velocity field is converging. Supernovae are implemented by injecting thermal energy into the 27 parent cells surrounding the sink particle and the total number of supernovae is calculated according to the size of the newly formed star cluster. Each supernova is assumed to produce 10$^{51}$ ergs of energy, which is transferred to the surrounding ISM with en efficiency $\eta_{SN}$=0.2. More details on the feedback implementation in RAMSES can be found in \citet{Dubois_Teyssier_2008}.

AMR is used here to capture the complex dynamics of the disk. In a box of 100 kpc, we use a coarse resolution of 256$^3$ with six levels of refinement, reaching an effective 4096$^3$ resolution in dense regions: An AMR level is activated if the mass in a cell exceeds 100 particles cm$^{-3}$, or if the local Jeans length is resolved with less than 10 cells.

\begin{table}[!th]
\caption{Summary of the \textit{Amalthea} models. The magnetic field strength refers to the initial value at center of the galaxy, in $\mu$G.}
  \begin{tabular*}{\linewidth}{@{\extracolsep{\fill}}llll}
    \hline
    {\textbf{Model}}  &  {\textbf{Feedback}}  & {\textbf{Max. \textbf{B}}}  & {\textbf{Max. resolution}} \\ 
    \hline
    \hline
     Amalthea M & yes & 0.1 &  20 pc  \\
     Amalthea Mnfb & no & 0.1 &  20 pc  \\
     Amalthea Mlr & yes & 0.1 & 40 pc   \\
     Amalthea W & yes & 0.01 & 20 pc   \\
    \hline
  \end{tabular*}
\label{tab:models}
\end{table}

%
\section{Dynamo action in the Amalthea models}
\label{sec:results}

\begin{figure}[h]
\begin{center}
\includegraphics[width=\linewidth]{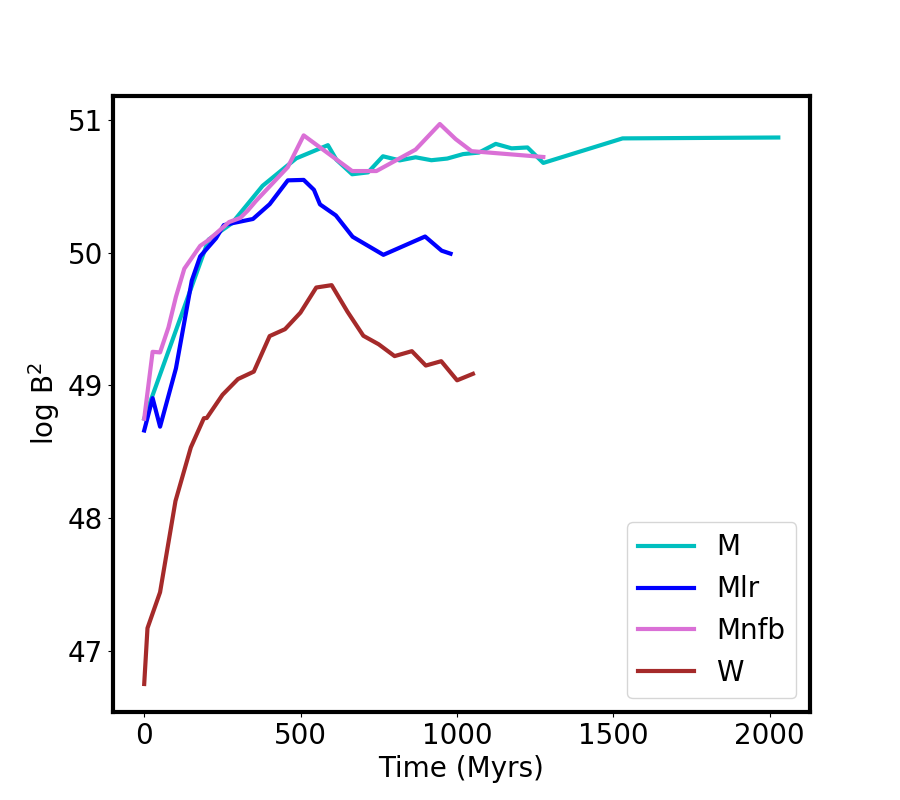}
\caption{Time evolution of the magnetic energy in a 25 kpc volume around Amalthea M, Mlr, Mnfb and W. The initial value corresponds to a mean field strength of about 0.003 $\mu$G in models M, Mlr, Mnfb and 3.e-4 $\mu$G in model W. The mean saturation value for models M and Mnfb is about 0.03~$\mu$G and 0.003~$\mu$G for model W.}
\label{field_amplification}
\end{center}
\end{figure}

Figure \ref{magnetic_evolution} shows how \emph{Amalthea M} and \emph{Amalthea Mnfb} evolve with time. The initially smooth exponential disk passes from a
relaxation phase that lasts about 200~Myrs. During this phase, stellar rings form at the inner regions of the~galaxy and pull the gas into similar formations. At about 0.6~Gyr, these inner structures turn into a bar, and two prominent spiral arms form in the outer disk. It is worth noting that this timescale is roughly the same as that for the magnetic field growth (Fig.\ref{field_amplification}). These structures continue evolving and at 1.5 Gyr, we observe that the central bar has shrunk significantly and is surrounded by a gaseous whirlpool.

The magnetic field follows this evolution closely. There is a field component within the disk that traces the rings at 0.18 Gyr and at later times, the bar and the spirals. At the same time, the small-scale turbulence in the disk, created by differential rotation, raises magnetic loops into the halo. This results in open magnetic field lines above and below the disk, and in the clear appearance of a poloidal component from an initially purely toroidal field. The magnetic field lines form vortices above and below the disk, indicative of a strong turbulent component.

The mean morphology of the field in Fig. \ref{magnetic_evolution} is reminiscent of kinematical mean-field dynamo model solutions \citep{Donner_1990}, specifically of the type generated by differential rotation in a thin-disk approximation.

It is natural then to look for an amplification of the magnetic energy with time. In Fig. \ref{field_amplification}, we show the magnetic field energy with time for
four models: the reference model M, model Mnfb, which is identical to M in every way, but does not include supernova feedback, model W, identical to M but with an initial field that is 10 times smaller, and model Mlr, which reaches a maximum resolution that is half that of model M. Models M and Mnfb follow nearly indistinguishable evolution paths, both reaching saturation at around 500 Myrs, with a total magnetic energy amplified by more than a factor of 100. The independence of the amplification from feedback is a strong indication that the amplification mechanism at work here depends on the large-scale dynamics of the galaxy and not on the supernova feedback.
The magnetic energy in model W is also amplified by roughly a factor of 100 on the same timescale of $\sim$500 Myrs and saturates at this new value.
Finally, model Mlr shows that the initial amplification of the field is converged. However, the saturation inevitably happens at a lower magnetic energy, due to the stronger numerical diffusivity of this model.


\subsection{Mean and turbulent components of the magnetic field}

In order to prove that what we are witnessing is, indeed, a mean-field dynamo, we need to compare its morphology to the available predictions from the dynamo theory.

For example, \citet{Stix_1975} showed that in a disk geometry a quadrupolar field should grow faster, while a dipolar field can arise as a result of asymmetries. \citet{Shukurov_2019} also presented a series of models  where quadrupolar fields are favored in the disk, while dipolar geometries arise in the halo.
In order to compare our findings to these analytical estimates, we need to split the magnetic field into a mean and a turbulent component.

We use a median filter to smooth the magnetic field at a given scale, which allows us to define the mean $\overline{\textbf{B}}$ and the turbulent component $\textbf{B}_{tur}$ of the magnetic field. 
In this method, the mean field is this smoothed component, while the turbulent component is the residual:

\begin{equation}
B_{tur} = \textbf{B}_{tot} - \overline{\textbf{B}}
\end{equation}
The filter is not used on the AMR data, but rather on quantities interpolated on a uniform grid of 48 pc resolution. 

\begin{figure*}[h!]
   \centering
     \subfloat[Model M, $\overline{\textbf{B}}$]{
       \includegraphics[width=0.45\linewidth]{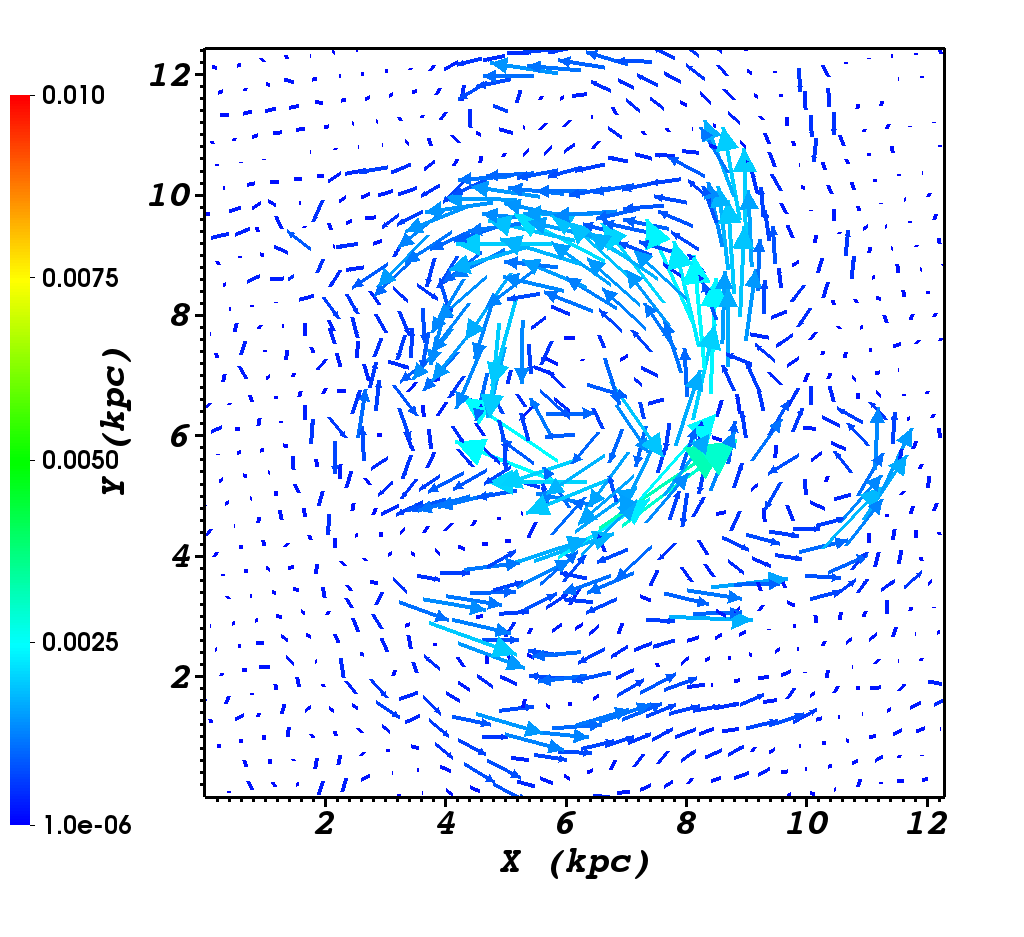}
     }
     \subfloat[Model M, $\textbf{B}_{tur}$]{
          \includegraphics[width=0.45\linewidth]{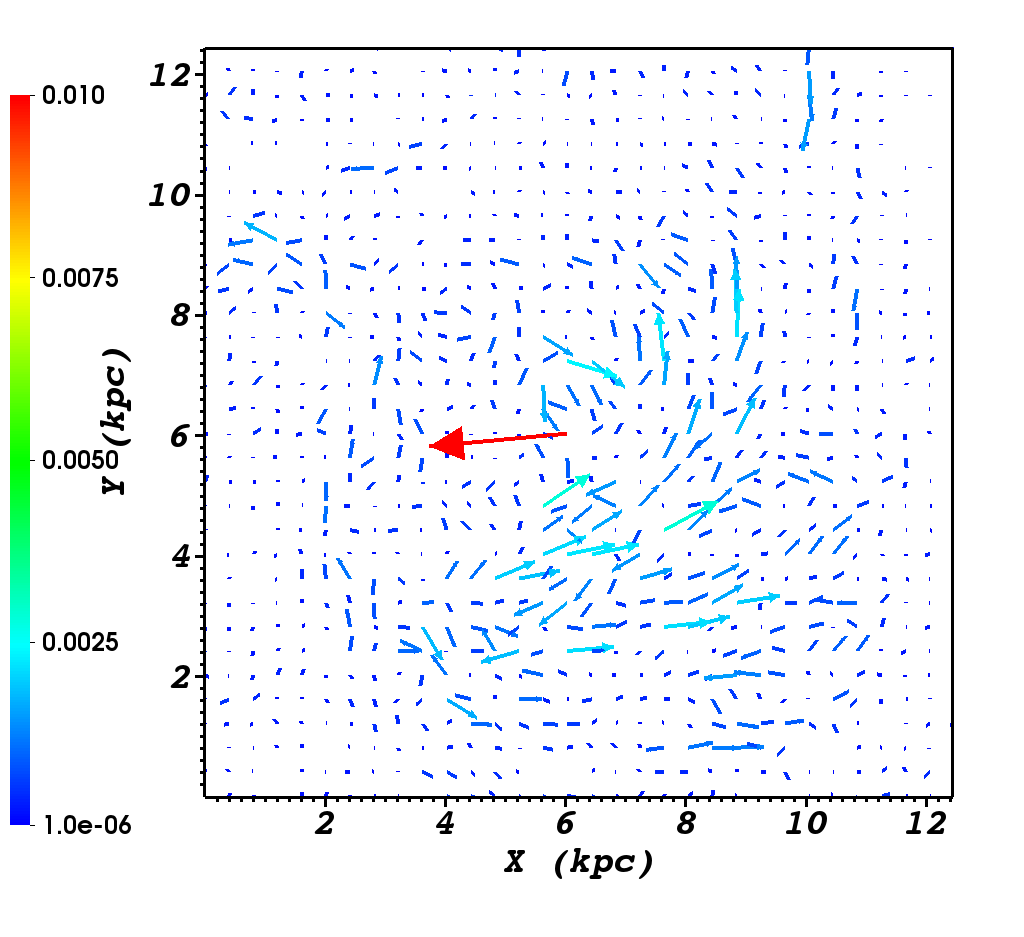}
     } \\ 
   \subfloat[Model Mnfb, $\overline{\textbf{B}}$]{
       \includegraphics[width=0.45\linewidth]{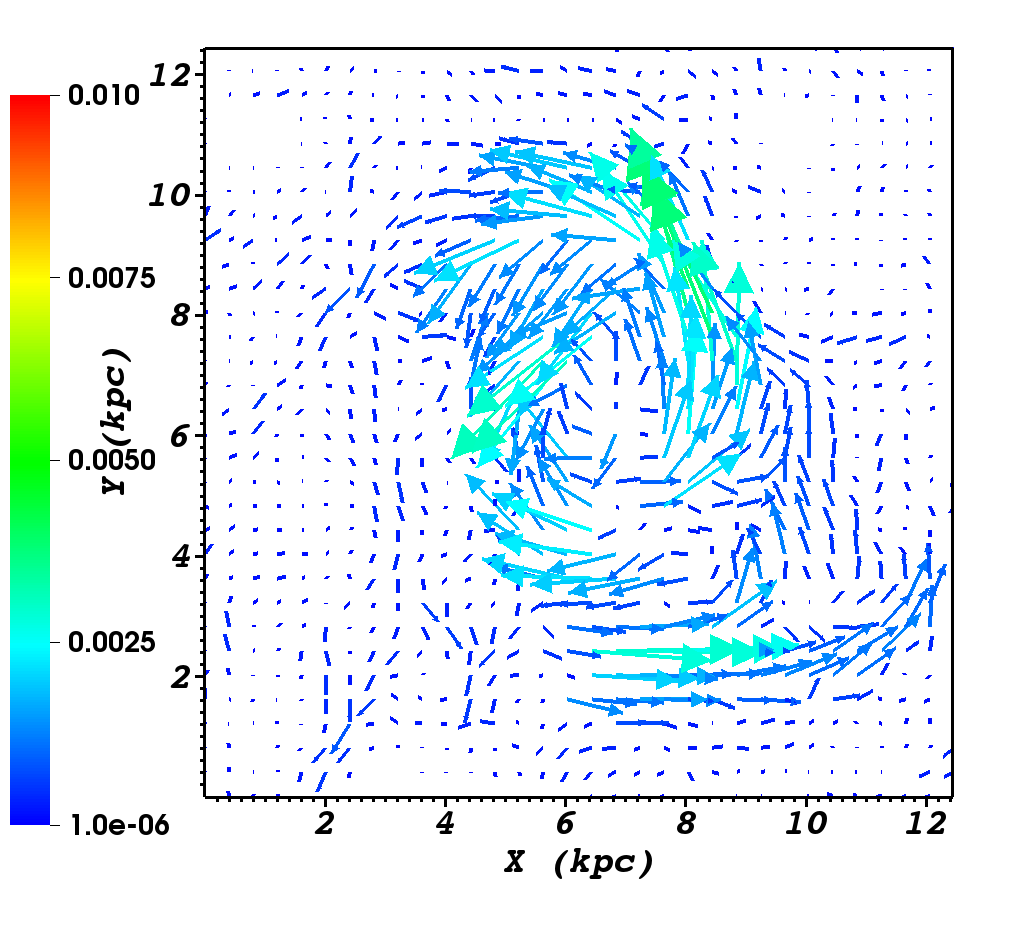}
     } 
   \subfloat[Model M, $\textbf{B}_{tur}$]{
       \includegraphics[width=0.45\linewidth]{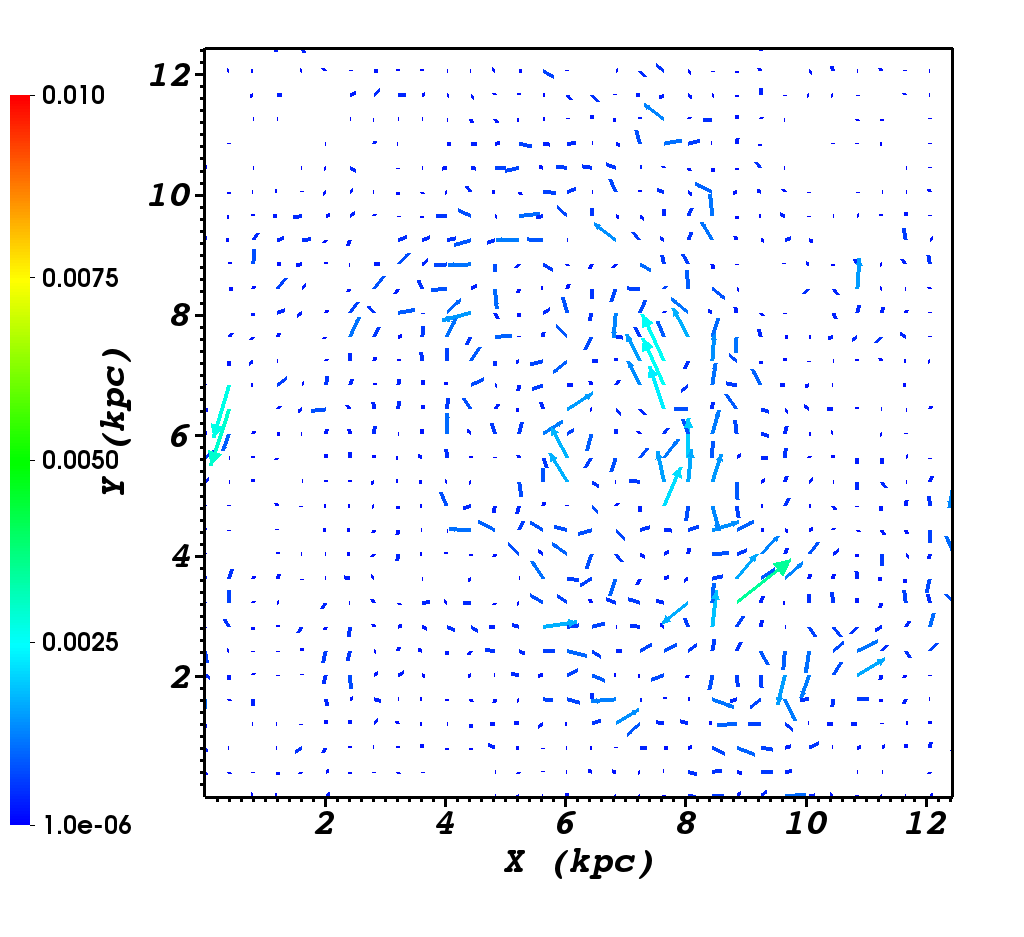}
    } \\  
   \caption{Mean (panels a and c) and turbulent (panels b and d) components of the magnetic field in Models M (panels a and b) and Mnfb (panels c and d), shown in a slice along the midplane of the galaxies, at 1275 Myrs into their evolution. The colors indicate the magnitude of the vectors drawn in each plot, in $\mu$G.}
     \label{mean_residual_b}
\end{figure*}

As an illustration, Fig. \ref{mean_residual_b} shows the mean and the turbulent components of the same galaxy segment in Amalthea M and Amalthea Mnfb after about
1.2 Gyr of evolution. It is worth noting the remarkable resemblance of the mean field component to observed magnetic fields of spiral galaxies, such as M51 \citep{Fletcher_2011}, or NGC
1068 \citep{Lopez-Rodriguez_2020}.

\begin{figure}[h]
\begin{center}
\includegraphics[width=\linewidth]{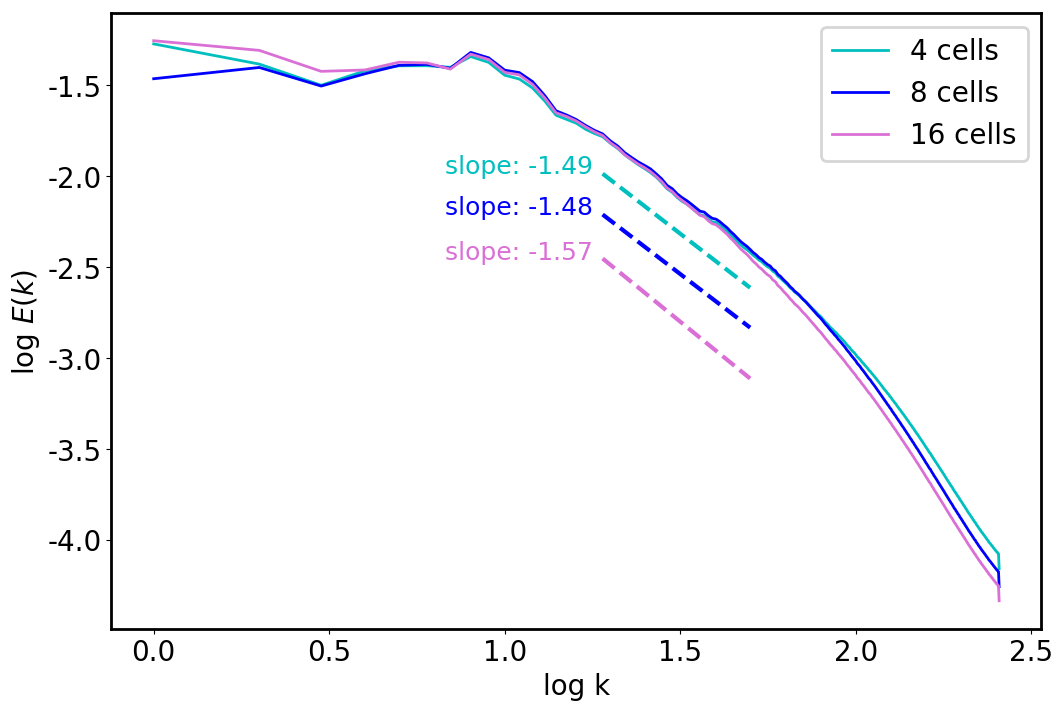}
\caption{Power spectra of the turbulent component of the magnetic field for different sizes of the averaging window. This example corresponds to Model M,
after 1Gyr of evolution, and a box of 25 kpc.}
\label{residual_pow_spec}
\end{center}
\end{figure}

The smoothing kernel used here measures 8x8x8 cells, which corresponds to a cubic region of about 390 pc size. 
The choice of this kernel size was made for two reasons: First, this physical scale corresponds to the thickness of the gaseous disk and to the size of superbubbles, which contribute to the creation of turbulence in the disk. Second, the power spectra of the residuals already appear converged at this kernel size (see Fig. \ref{residual_pow_spec}).

Figure \ref{residual_pow_spec} contains the power spectra of the residual (turbulent) component of the magnetic field using different sizes of the smoothing
window. They all produce power-law slopes~consistent with Kolmogorov turbulence in the inertial range, and have a shape very similar to the power spectra of isotropic, helical turbulence \citep{Brandenburg_2012}.

\begin{figure*}[h!]
   \centering
     \subfloat[Model M, $B_x$ projection]{
      \includegraphics[trim=30 50 50 10,clip,width=0.45\linewidth]{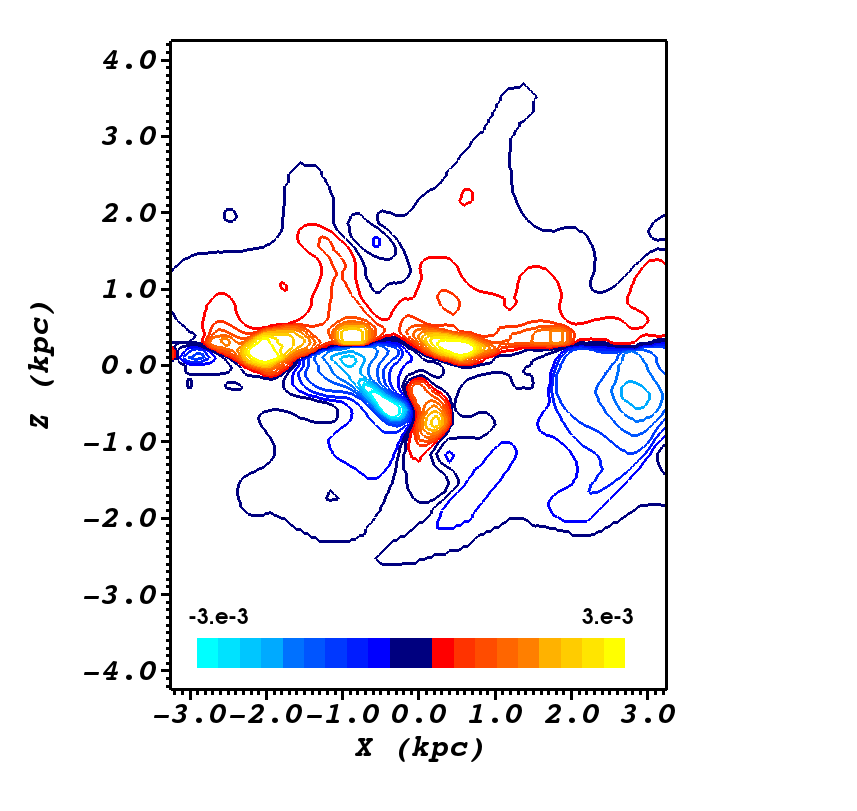}
     }
     \subfloat[Model M, $B_y$ projection]{
     \includegraphics[trim=30 50 50 10,clip,width=0.45\linewidth]{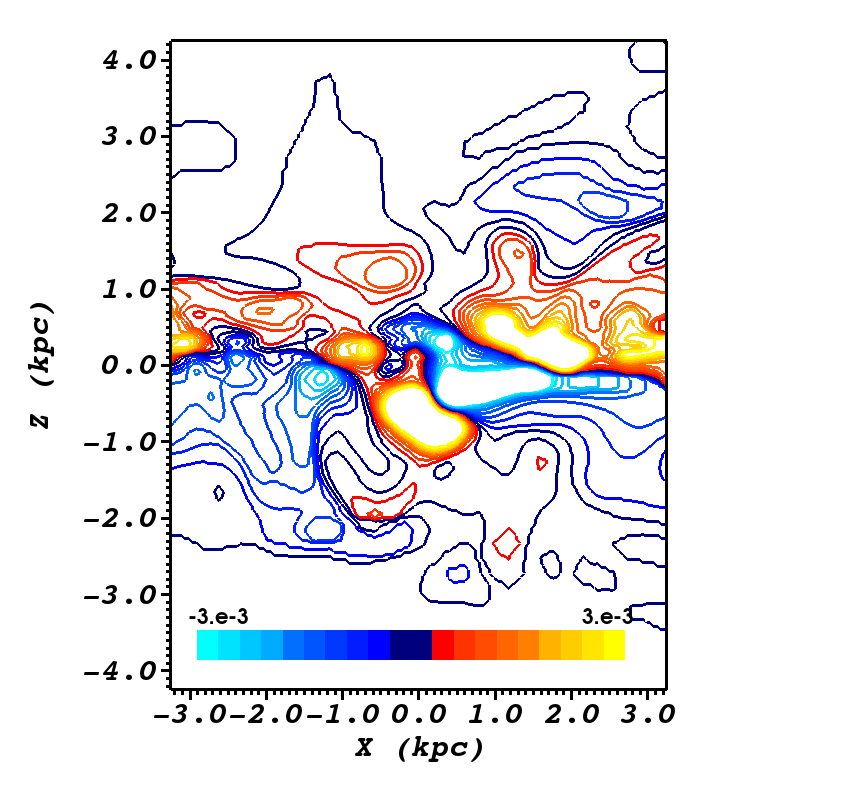}
     } \\ 
   \subfloat[Model Mnfb, $B_x$ projection]{
       \includegraphics[trim=30 50 50 10,clip,width=0.45\linewidth]{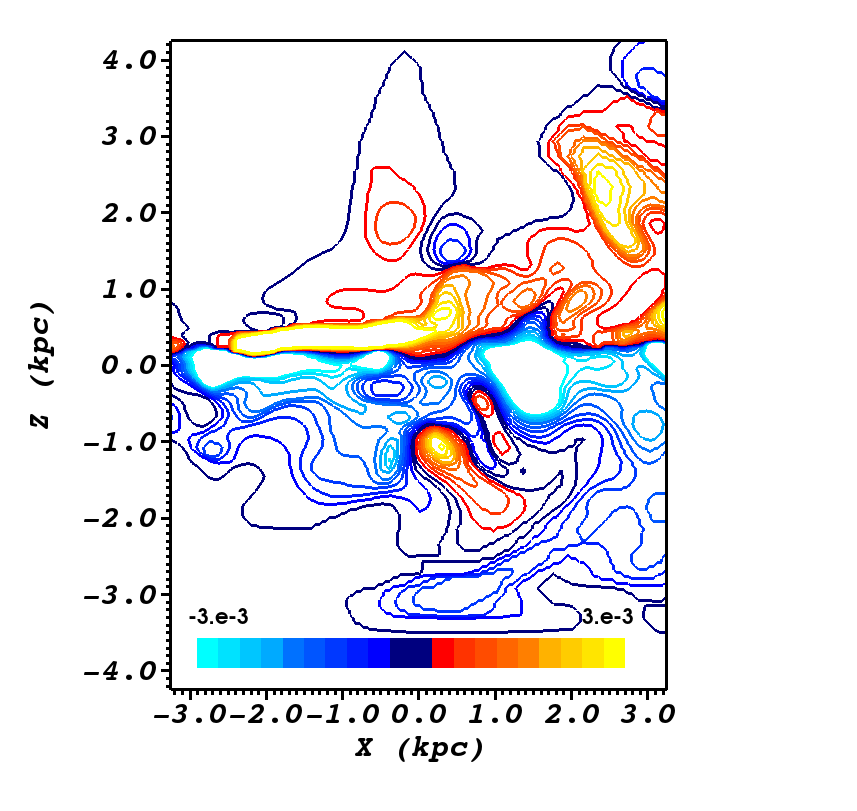}
     } 
   \subfloat[Model Mnfb, $B_y$ projection]{
        \includegraphics[trim=30 50 50 10,clip,width=0.45\linewidth]{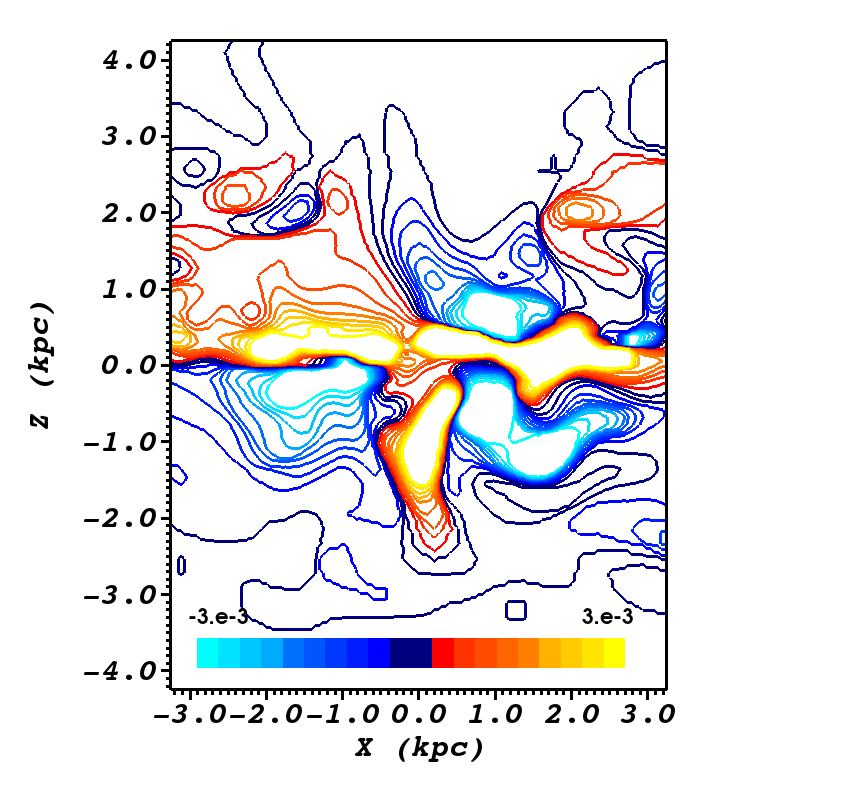}
    } \\  
   \caption{Contour plots of the Bx-component on the x=0 plane (panels a and c) and the By-component on the y=0 plane (panels b and d) of the mean field for the Amalthea M (panels a and b) and Mnfb (panels c and d) simulations. Essentially, the plots represent local cuts of the toroidal component of the field.}
     \label{contours_mag_field_size8}
\end{figure*}

\begin{figure*}[h!]
   \centering
     \subfloat[Model M, $B_x$ projection]{
       \includegraphics[trim=50 50 0 200,clip,width=0.45\linewidth]{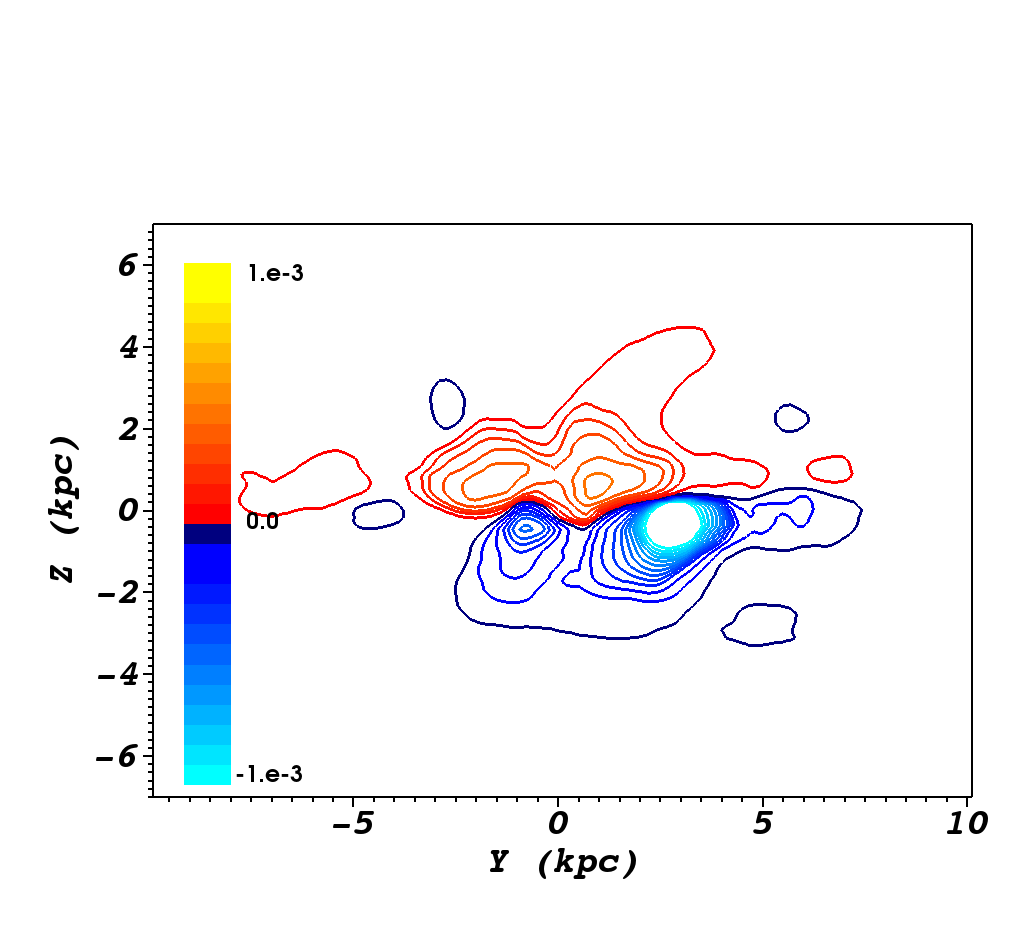}
     }
     \subfloat[Model M, $B_y$ projection]{
          \includegraphics[trim=50 50 0 150,clip,width=0.45\linewidth]{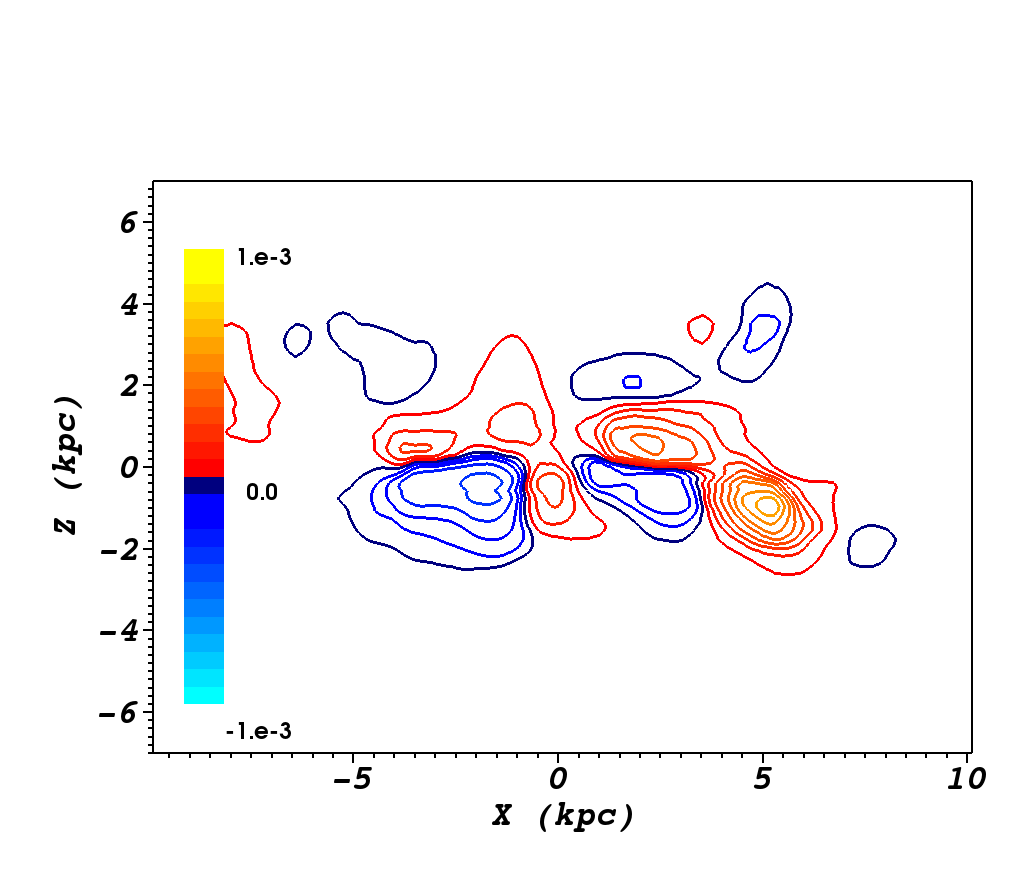}
     } \\ 
   \subfloat[Model Mnfb, $B_x$ projection]{
       \includegraphics[trim=50 50 0 200,clip,width=0.45\linewidth]{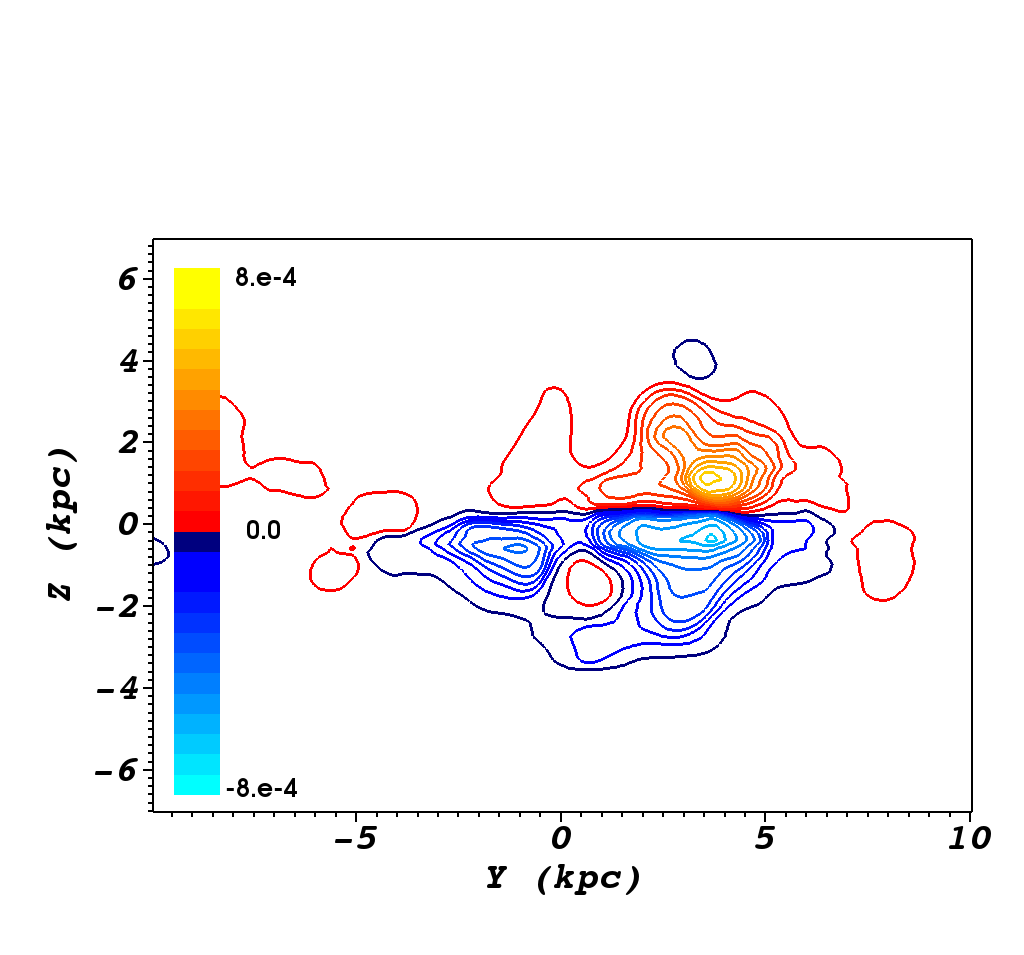}
     } 
   \subfloat[Model M, $B_y$ projection]{
       \includegraphics[trim=50 50 0 200,clip,width=0.45\linewidth]{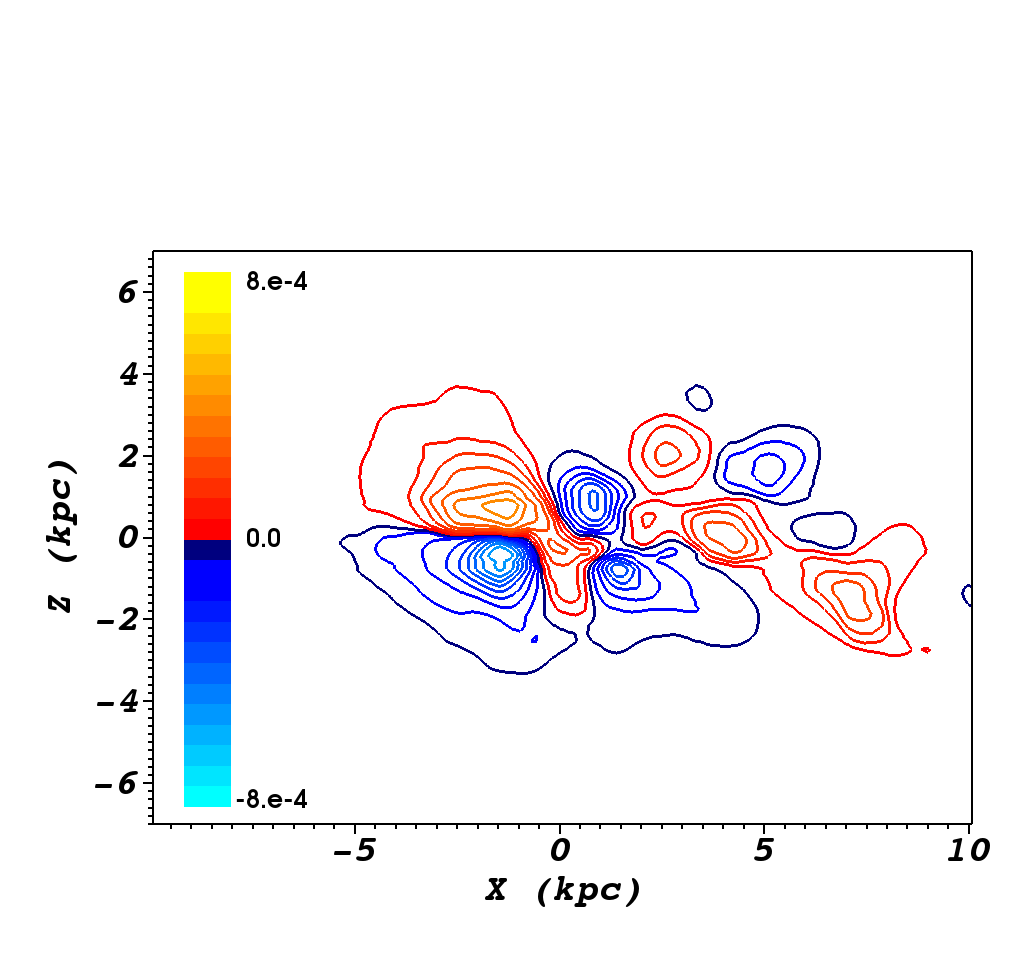}
    } \\  
   \caption{As in Fig.\ref{contours_mag_field_size8}, but using a 16x16x16 cell smoothing window. The quadrupolar shape of the field is more evident here, and in complete agreement with analytical dynamo models.}
     \label{contours_mag_field_size16}
\end{figure*}

In Fig. \ref{contours_mag_field_size8} we show cuts of the mean Bx and By components of the field on the x=0 and y=0 planes, respectively. The top panel corresponds to model Mnfb, and the bottom panel to model M. Both models develop a dipolar mean field in the halo, and a quadrupolar field closer to the disk, although the shapes are better defined in model M. These results were produced using the 8 cell window. 
The mean field using the 16 cell kernel window is shown in Fig. \ref{contours_mag_field_size16}). In that case, the dipolar/quadrupolar shape of the mean field is more evident, and is almost identical to the findings of \citet{Stix_1975}.

\subsection{Kinetic and current helicity evolution}

The operation of the galactic dynamo depends on the generation of helical turbulence in the disk \citep{Parker_1955}, which amplifies the magnetic field through a process also known as the $\alpha$-effect \citep{SKR_1966}.

Assuming that both the magnetic field and the velocity can be decomposed into a mean and a turbulent component: $\textbf{B}=\overline{\textbf{B}}+\textbf{B}_{tur}$, $\textbf{v}=\overline{\textbf{v}}+\textbf{v}_{tur}$, and that the turbulent component is isotropic, we can take the average of Eq. \ref{induction} to obtain:
\begin{equation}
    \frac{\partial\overline{\bf{B}}}{\partial t} -\nabla\times(\overline{\bf{v}}\times\overline{\bf{B}}+\epsilon) = 0 
    \label{mean_dynamo_eq}
\end{equation}
where $\epsilon$ is the electromotive force due to turbulent motions, 
$\epsilon=\overline{\textbf{v}_{tur}\times \textbf{B}_{tur}}$.
It is worth noting here that the full dynamo equation also includes the magnetic diffusivity $\eta_{M}$ but, in these simulations, we do not explicitly include magnetic diffusion terms. Here, magnetic diffusion happens only due to the finite resolution of the grid. 

An assumption of the mean-field dynamo theory is that $\overline{\textbf{v}_{tur}\times \textbf{B}_{tur}}=\alpha\overline{\textbf{B}}-\eta_T\nabla\times\overline{\textbf{B}}$, where $\alpha$ and $\eta_T$ are the $\alpha$-effect and the turbulent magnetic diffusivity tensors. In particular, at the high conductivity limit for isotropic turbulence, and ignoring the effect of Lorentz forces, the $\alpha$-effect tensor can be written as: \citep{Moffatt_1978,Krause_1980}:
\begin{equation}
    \alpha_K = -\frac{\tau}{3}\cdot (\overline{\textbf{v}_{tur}\cdot(\nabla\times\textbf{v}_{tur}}))
    \label{alpha_tensor}
\end{equation}
where $\tau$ is the correlation time of the turbulence. The tensor $\alpha_K$ is proportional to K$_{tur}$, the kinetic helicity of the fluid due to turbulent motions:

\begin{equation}
K_{tur}=\int \textbf{v}_{tur}\cdot(\nabla\times\textbf{v}_{tur})\,dV.
\label{eq:kturb}
\end{equation}

When Lorentz forces become significant, there is a second contribution to the $\alpha$-tensor, which can be written as \citep[e.g.][]{Brandenburg_2005}:
\begin{equation}
    \alpha_M = -\frac{\tau}{3}\cdot \frac{(\overline{\textbf{B}_{tur}\cdot(\nabla\times\textbf{B}_{tur}})}{4\pi\rho}.
    \label{eq:cturb}
\end{equation}
$\alpha_M$ is proportional to $C_{tur}$, the small-scale current helicity due to the turbulent component of the magnetic field: 
\begin{equation}
C_{tur}=\int \textbf{B}_{tur}\cdot(\nabla\times\textbf{B}_{tur})\,dV
\label{eq:hturb}
\end{equation}
The full $\alpha$-tensor is therefore the sum of the kinetic and magnetic terms, $\alpha$=$\alpha_K$ + $\alpha_M$.

Calculating the $\alpha$ and $\eta_M$ tensors is an involved process, and a subject of many numerical and analytical works \citep[e.g][]{Ferriere_1998, Hubbard_2009, Rheinhardt_2010}. Here, we will limit ourselves to showing that the models host helical turbulence, which is a necessary ingredient of the $\alpha$ dynamo process.

In the \emph{Amalthea} simulations, turbulence is generated by the differential rotation of the galaxy and by the supernova explosions (in all models apart from Mnfb, which does not include supernova feedback). We can obtain an estimate for the root mean square (rms) turbulent velocity in the different models by separating the mean from the turbulent velocity components using the same procedure as for the magnetic field. This yields $\textbf{v}_{rms}\simeq 3-14$ km/sec in the different models at different evolution times. As expected, in the simulations that include feedback, the turbulent velocities fluctuate on the typical timescales of star formation cycles, $\sim$30-50 Myrs.

\begin{figure*}[h!]
   \centering
     \subfloat[Model M, K$_{tot}$]{
       \includegraphics[trim=50 50 0 0, width=0.45\linewidth]{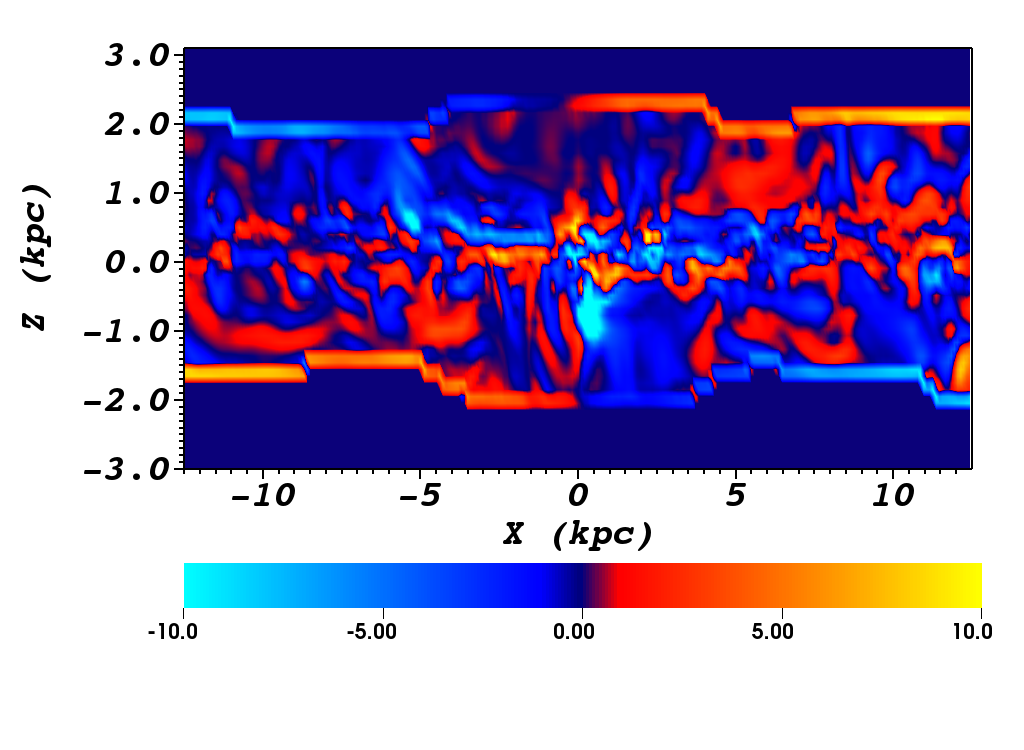}
     }
     \subfloat[Model M, K$_{tur}$]{
          \includegraphics[trim=50 50 0 0, width=0.45\linewidth]{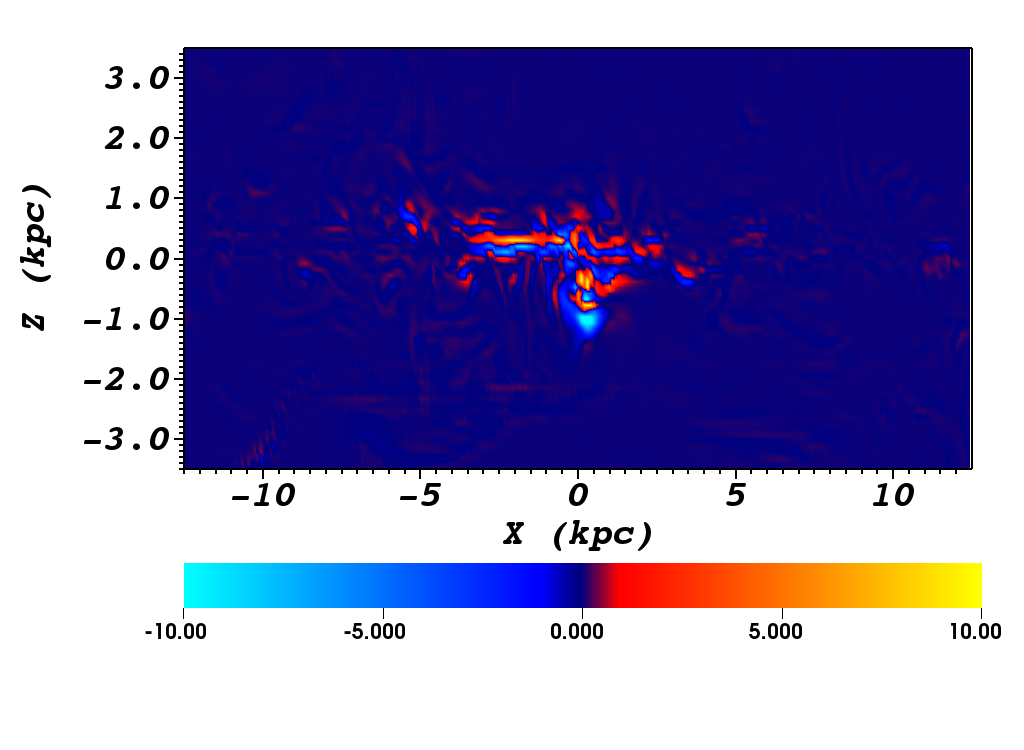}
     } \\ 
   \subfloat[Model Mnfb, K$_{tot}$]{
       \includegraphics[trim=50 50 0 0,width=0.45\linewidth]{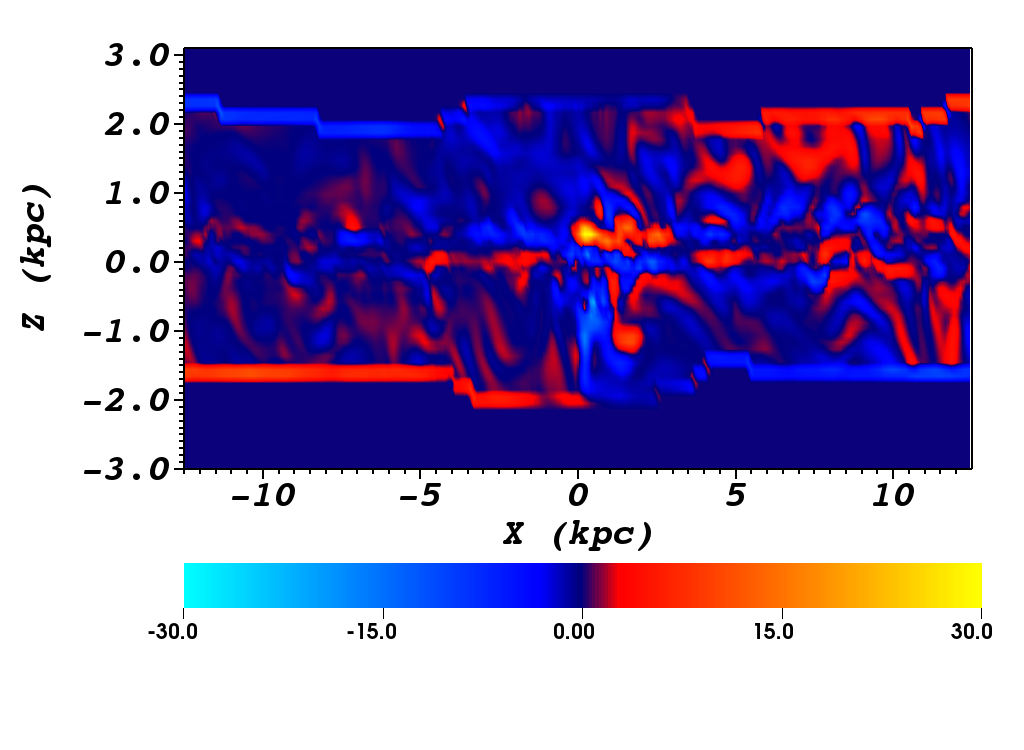}
     } 
   \subfloat[Model Mnfb, K$_{tur}$]{
       \includegraphics[trim=70 30 0 0,width=0.45\linewidth]{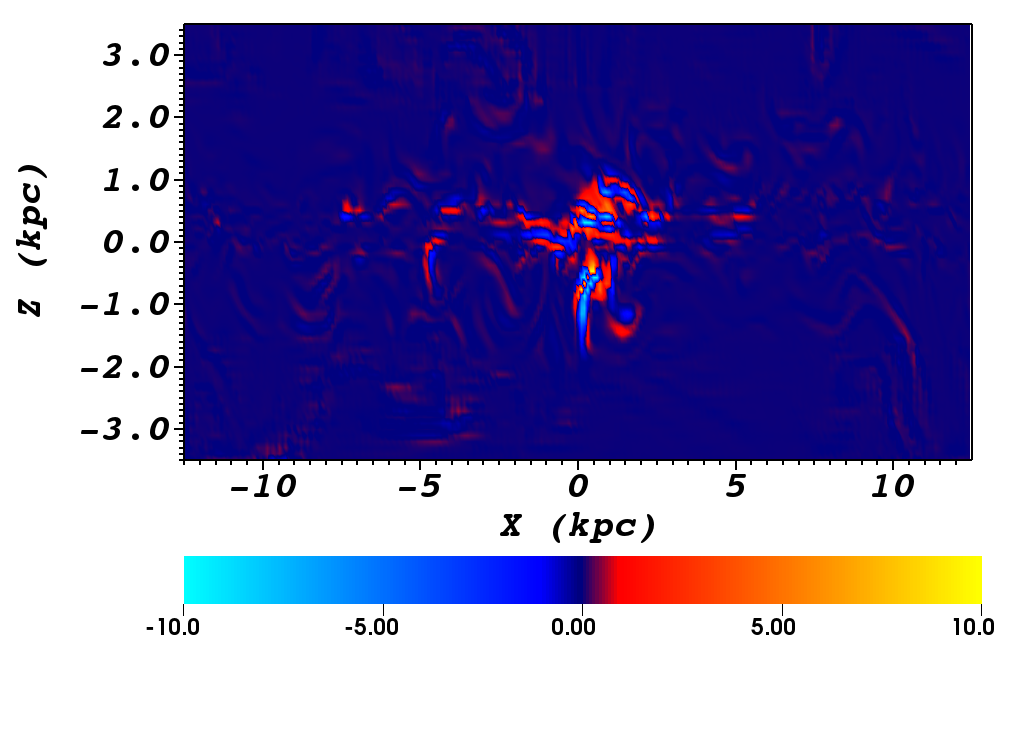}    } \\  

   \caption{Slices of the total (panels a and c) and turbulent (panels b and d) components of the kinetic helicity in Models M (panels a and b) and Mnfb (panels c and d), along the xz midplane of the galaxies, at 1275 Myrs into their evolution. The quantities are shown in code units, and the values were saturated at the range shown in order to enhance contrast.}
     \label{kinetic_helicity_slices}
\end{figure*}

Figure \ref{kinetic_helicity_slices} shows slices of the total and the turbulent kinetic helicities in models M and Mnfb for the same snapshot at 1275 Myrs. It is clear that helical turbulence is generated in both simulations, and that the kinetic helicity is antisymmetric across the galactic midplane.

\begin{figure*}
    \centering
    
    \subfloat[K$_{tot}$]{
       \includegraphics[trim=0 0 70 50, width=0.45\linewidth]{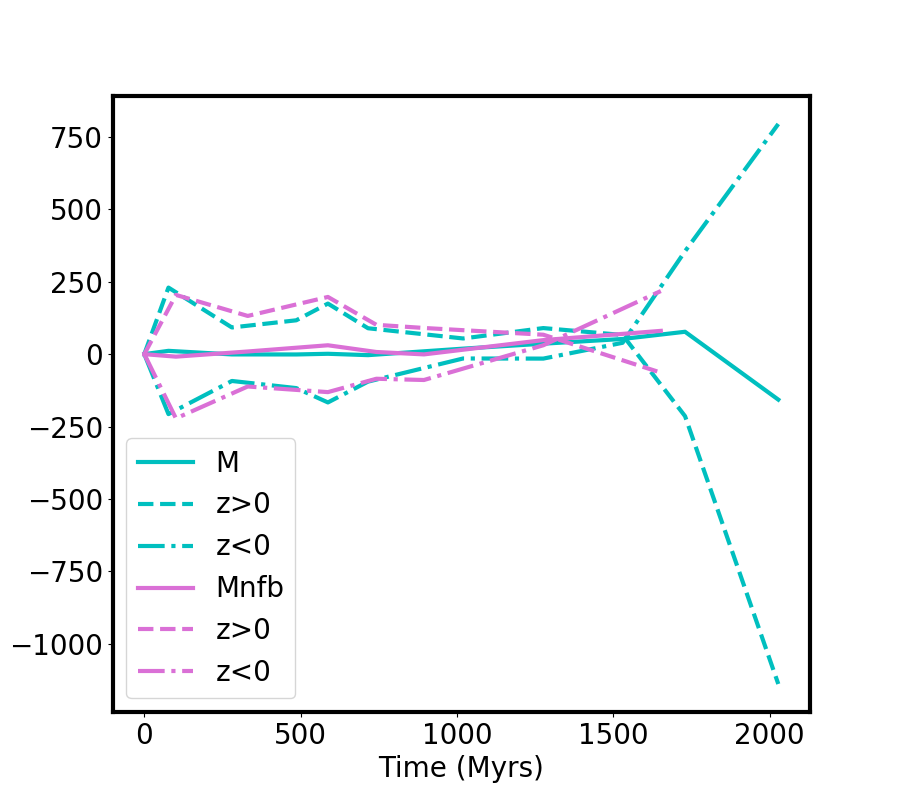}
     }
     \subfloat[K$_{tur}$]{
          \includegraphics[trim=0 0 70 50, width=0.45\linewidth]{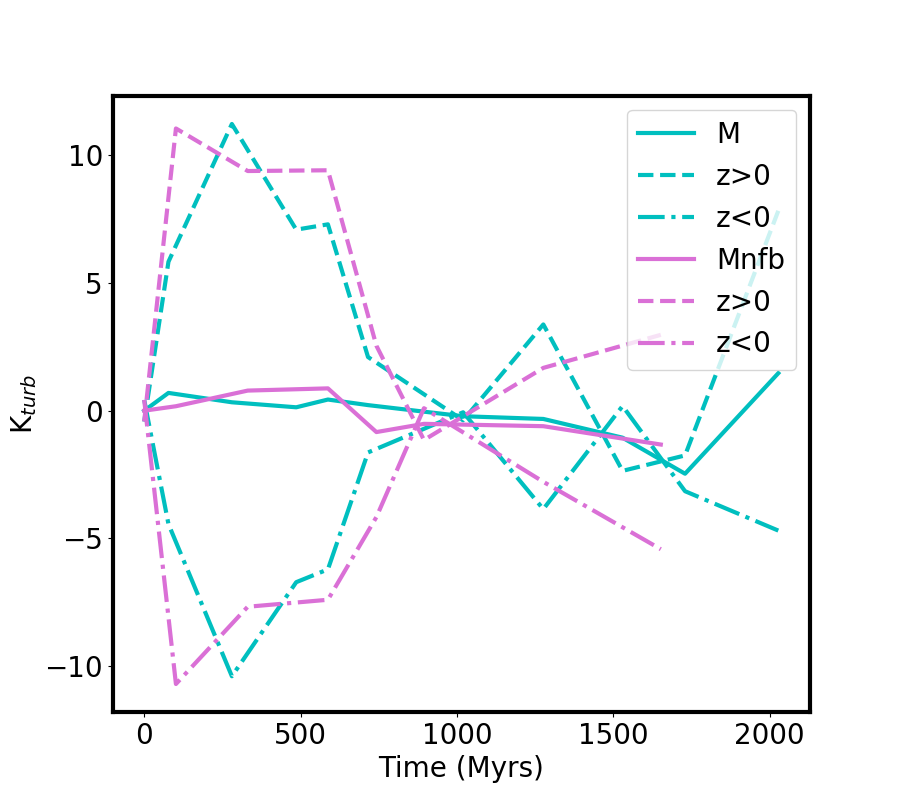}
     } \\ 
   \subfloat[C$_{tot}$]{
       \includegraphics[trim=0 0 70 50,width=0.45\linewidth]{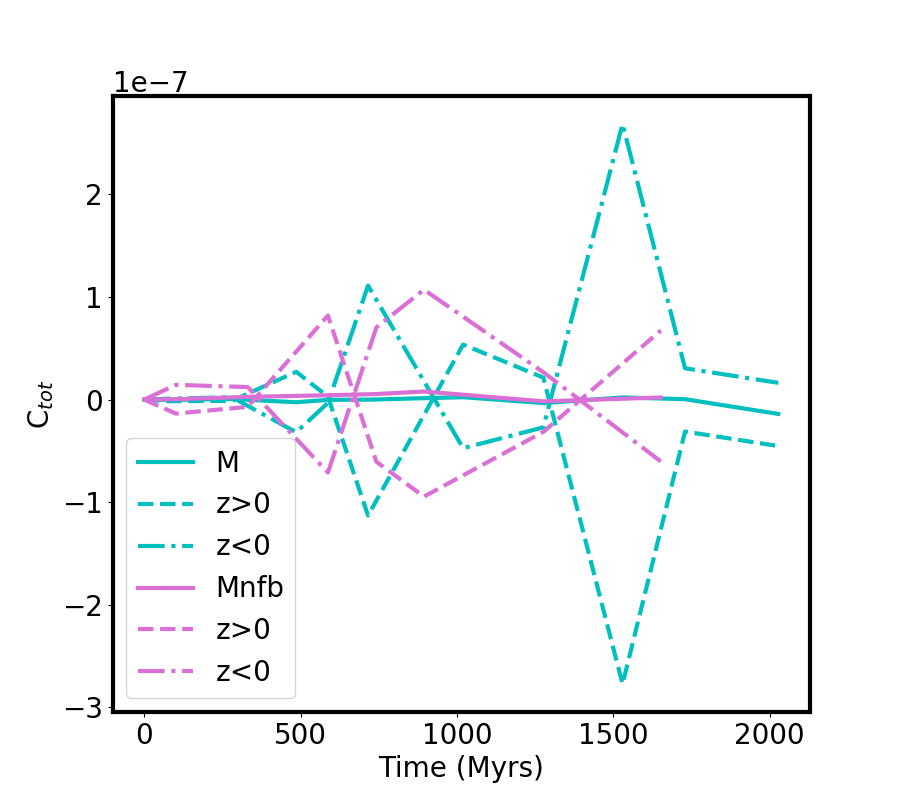}
     } 
   \subfloat[C$_{tur}$]{
       \includegraphics[trim=0 0 70 50,width=0.45\linewidth]{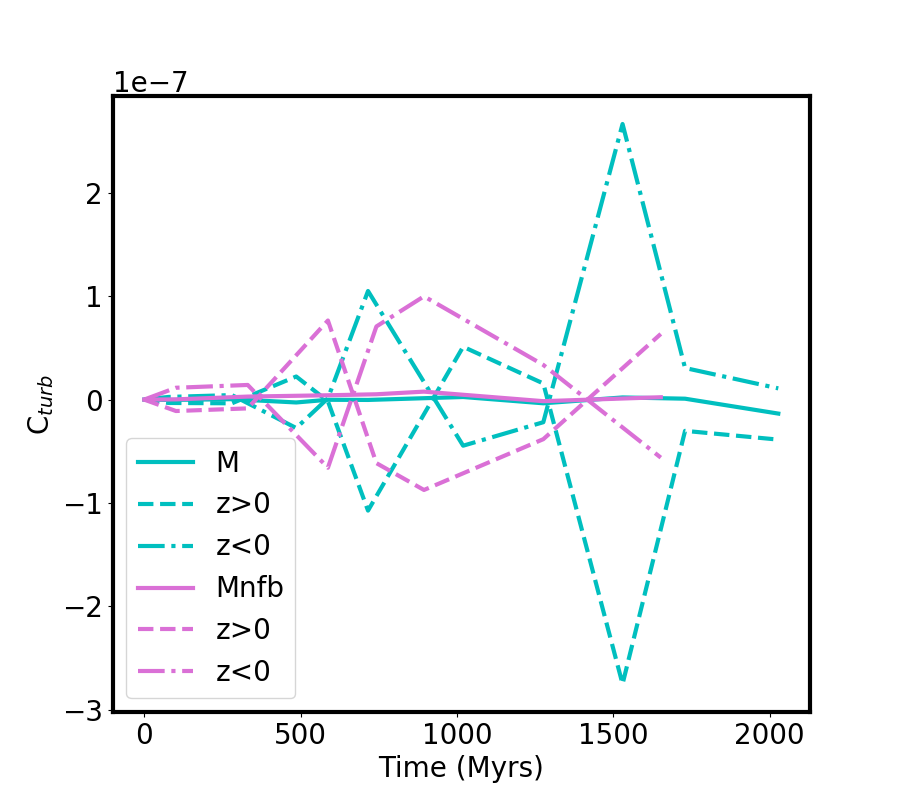}} \\  
    
    \caption{Total (left) and turbulent (right) kinetic (Eq. \ref{eq:kturb}) and current helicities (Eq. \ref{eq:cturb}) as a function of time in models M and Mnfb. Kinetic helicities are in km$^2$~sec$^{-2}$~kpc$^{-1}$ and current helicities in $\mu G^2$~kpc$^{-1}$ The dashed and dotted-dashed lines show the same quantities in the upper and lower parts of the computational volume, respectively. Notice that the total and the turbulent current helicities are essentially identical.}
    \label{kinetic_helicity_time}
\end{figure*}

The top panel of Fig. \ref{kinetic_helicity_time} shows the evolution of the total and the turbulent kinetic helicities with time for the same two models. In addition to the total volume integrals, we are showing the kinetic helicities in the upper and the lower parts of the box separately. Here, z=0 is defined as the z-coordinate of the center of mass of the gas, and the volume integrals are taken over a 25~kpc$^3$ box centered at the center of mass. 
Both the total and the turbulent kinetic helicity change sign across the galactic midplane, while the helicities integrated over the entire volume are always very close to zero. This antisymmetry indicates that the turbulence in the disk is helical throughout the evolution of the models, and an $\alpha$ mechanism is active in these simulations.

The evolution of the total and turbulent current helicities with time is shown in the bottom panel of Fig. \ref{kinetic_helicity_time}. The total current helicity and the turbulent current helicity are identical, meaning that the current helicity in the volume is generated entirely on the smallest scales. There are multiple sign reversals across the midplane, with the total current helicity in the volume always close to zero. These strong local fluctuations indicate that the current helicity has not reached saturation.

\begin{figure*}[h]
\begin{center}
\includegraphics[width=0.8\linewidth]{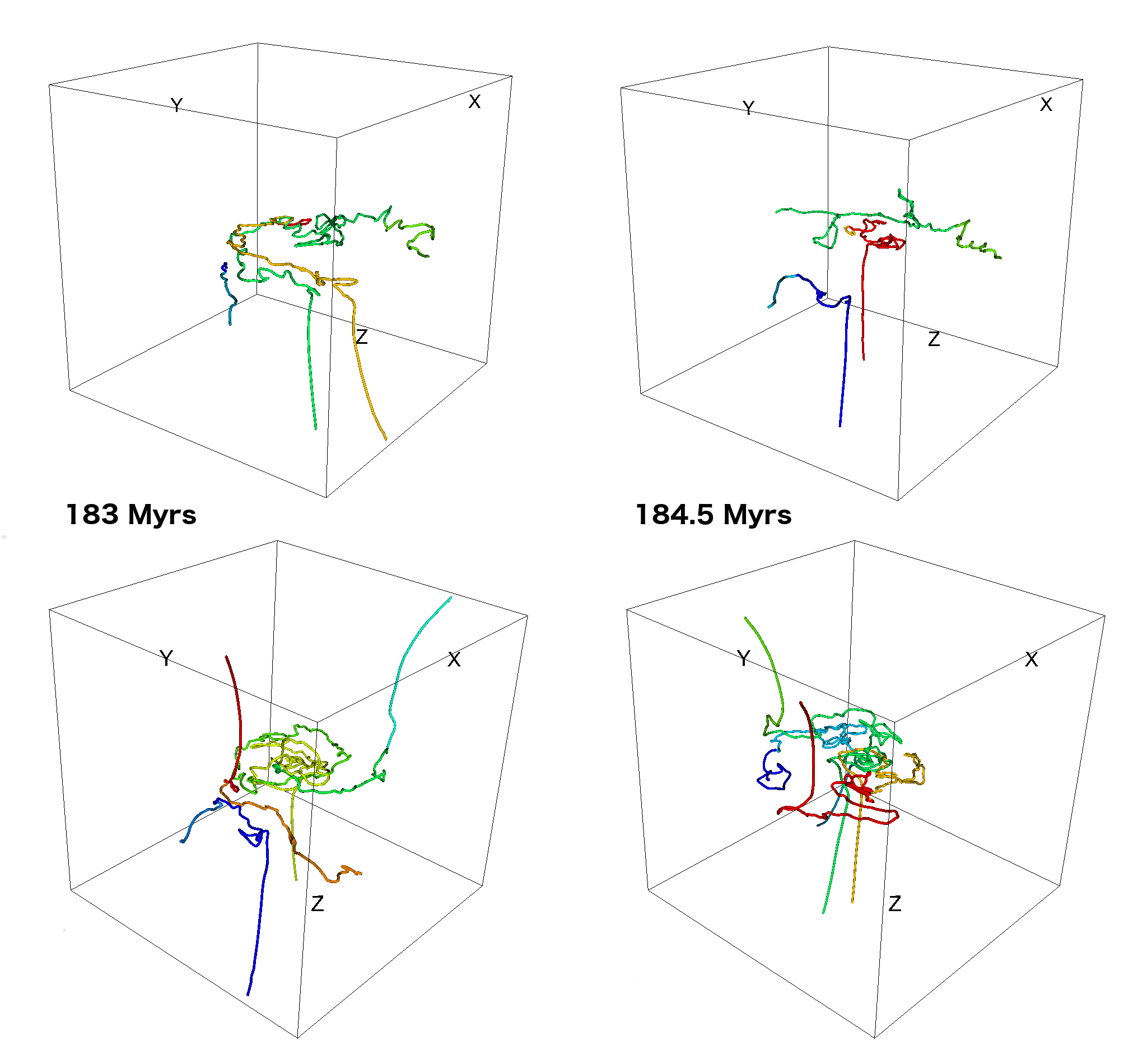}
\caption{Illustration of the $\alpha$-and the $\omega$-dynamo mechanisms acting in the simulation. The top row shows field lines rising, rotating in opposite directions above and below the disk, and reconnecting to create poloidal field from the toroidal component ($\alpha$-dynamo). The bottom row shows field lines winding up by differential rotation to create toroidal field from the poloidal ($\omega$-dynamo). The box size in this illustration is 24~kpc.}
\label{field_lines}
\end{center}
\end{figure*}

\subsection{Magnetic field lines}

Witnessing how magnetic field lines re-order is an excellent diagnostic for identifying the amplification mechanism. This reshaping process is discernible in Fig. \ref{field_amplification}, but more clearly illustrated in Fig. \ref{field_lines}, where we have drawn four individual magnetic field lines from model \emph{Amalthea~M} as they evolve from one code output to the next, 1.5 Myrs later. The starting points for the line integration are chosen randomly for the first snapshot, and kept the same in the following snapshots. The integration is done using the "Streamline" function of the ViSit visualization code \citep{HPV:VisIt} with a fourth-order Runge-Kutta scheme.

It is clear from these visualizations that the reconnection rate is faster than the 1.5 Myrs cadence between snapshots. A clear example is the yellow field line in the top panel, which is reduced to a small loop after 1.5 Myrs.

The top row of the figure shows the $\alpha$-dynamo mechanism in action: magnetic loops rise above and below the disk, spinning due to the differential rotation of the disk. At the same time, they constantly reconnect, creating the poloidal component and open field lines. The bottom row of the same figure illustrates how the $\omega$-dynamo mechanism operates. Vertical field lines are tightly wound up by the disk rotation, and eventually reconnect to form closed loops parallel to the disk.

Thus, in complete agreement with dynamo theory, the initially toroidal field develops a poloidal component, which in turn replenishes the toroidal component. However, the magnetic energy in Amalthea M never surpasses 0.1 percent of the turbulent kinetic energy, and remains a million times smaller than the total kinetic energy of the galaxy. This result is particularly important when considering the long-term evolution of the galaxy and its magnetic field: It demonstrates that a
mean-field dynamo naturally arises from galactic dynamics alone, and that it amplifies the initially negligible magnetic field by at least one order of magnitude.

\section{Discussion and conclusions}
\label{sec:discussion}

In this work, we have presented global, direct numerical simulations of spiral galaxies with and without supernova feedback.
In all models, we have found the characteristic features of $\alpha$-$\omega$ dynamo amplification: helical turbulence, reconnecting magnetic field lines, and a final magnetic field with a quadrupolar topology.

The initial magnetic energy in all these simulations is amplified by roughly a factor of 100 within about 500 Myrs, even when supernova feedback is absent. 
These results indicate that an $\alpha$-$\omega$ dynamo is a natural consequence of the dynamics of a massive spiral galaxy and that its action is independent of supernova feedback.

In the classical mean-field dynamo theory, the magnetic field continues to grow until it reaches saturation at rough equipartition with the turbulent kinetic energy of the gas.
In our models, however, the magnetic energy never even approaches equipartition. The simulations with the strongest initial magnetization (models M, Mnfb and Mlr) show saturation when the magnetic energy is roughly a millionth of the kinetic energy.
There are two possible explanations for this behavior: either there is a back-reaction of the magnetic field on the velocity flow that hinders further growth, or the amplification mechanism itself is not efficient after 500 Myrs.

In the first case, we would be witnessing the "alpha-quenching" mechanism. According to \citet{Vainshtein_Cattaneo_1992}, the growth of a small-scale random magnetic field can halt the growth of the galactic magnetic field, as Lorentz forces inhibit the formation of new turbulent diffusive structures. Indeed, \citet{EN_2018} showed that, in a setup very similar to \emph{Amalthea}, a random magnetic field starts to develop in less than 25 Myrs.  Therefore, it is entirely possible that the Vainstein \& Cattaneo quenching will affect the \emph{Amalthea} models at some point of their evolution.
Vainstein \& Cattaneo also calculated that the maximum expected growth of the large-scale magnetic field by a dynamo, can be of the order of $R_m^{1/2}$ smaller than equipartition, where $R_m$ is the magnetic Reynolds number of the flow. For the Galaxy, this number is of the order of 10$^7$. 
However, this problem can be alleviated, for example, by a galactic fountain flow \citep{Shukurov_2006}, or even a galactic "breeze" \citep{DS_2013} that removes small scale magnetic helicity from the disk and allows the dynamo mechanism to continue operating.
Neither of these phenomena is present in the \emph{Amalthea} models. A longer evolution of the simulations will show if this occurs later in the life of a galaxy, or if the galactic environment can play a role in the magnetic field growth.

However, there are two reasons to believe that we are not witnessing a saturation due to alpha quenching. The first one is that the local values of the current helicity in the simulations have not reached saturation. Instead, they fluctuates around zero, changing sign constantly across the galactic midplane.

The second reason is that in the \emph{Amalthea} models, the saturation value of the magnetic field seems to depend on its initial strength: Models M and W are identical in every other way apart from their initial magnetic energy, which is a hundred times lower in model W. Yet the amplification of the field in model W does not continue after reaching the initial value of model M. Instead, it saturates at 500 Myrs, indicating that the models share a common amplification factor and a common growth rate and not a common final magnetic field strength. In other words, the initial difference in energy between the models is preserved.

This leads us to the second possibility: that the amplification mechanism weakens after $\sim$500 Myrs. This hypothesis is supported by the evolution of the kinetic helicity, which is a measure of the alpha dynamo action. After about 500 Myrs, the degree of asymmetry of the turbulent kinetic helicity decreases, which implies that the generation of helical turbulence slows down and the magnetic field amplification cannot be as efficient. 
This timescale coincides with the formation of large-scale dynamical features, such as spiral arms and a central bar, which are strong drivers of turbulence.
However, it is not possible to draw general conclusions based on these models alone. In order to investigate this hypothesis further, we need to explore a vast parameter space, which includes the initial morphology and strength of the magnetic field, as well as the dynamical evolution of the galaxy. These simulations require a large computational effort and will be the subject of a follow-up study.

Still, the saturation values of the magnetic field in all our models are at least an order of magnitude below the values typically observed in our Galaxy. This disagreement underlines the necessity for additional amplification mechanisms. For instance, the inclusion of cosmic rays could affect the buoyancy of the gas or drive a galactic wind. Both these effects could lead to a faster amplification of the field.

The cosmological growth of the Galaxy itself may be another strong candidate for the necessary boost in field amplification. In cosmological simulations that include magnetic fields, an amplification up to 10 percent of equipartition values is observed already at redshift 2-3 \citep{Pakmor_2017}. This rapid growth of the field is in apparent disagreement with the argument that wants gravitational amplification to be inefficient due to diffusion. The key probably lies in a continuous gravitational growth that replenishes the magnetic field faster than diffusion can transform it.

Then the question of how the cosmological growth of a galaxy affects the evolution of a dynamo becomes essential for understanding the origin of galactic magnetic fields. It indicates the future direction of galaxy evolution models: the inclusion of magnetization at a higher resolution.

\emph{Acknowledgments}

We are grateful to Vincent Pelgrims for observations that helped refine our calculations of the helicity, to Gustavo Guerrero for useful discussions on the dynamo saturation mechanism, and to the anonymous referee for suggestions that greatly improved this manuscript.
This research has received funding from a Marie Curie Action of the European Union (Grant agreement number 749073) and from the European
Research Council under the European Union's Horizon 2020 research and Innovation program, under grant agreements No 771282 and 740120. The largest part of
these models were carried out at the IRENE supercomputer, as part of the PRACE Project 2018194718. Part of this work was performed under the Project HPC-EUROPA3 (INFRAIA-2016-1-730897), with the support of the EC Research Innovation Action under the H2020 Program. In particular, E.N. gratefully acknowledges the support of the Max-Planck for Astrophysics and the computer resources and technical support provided by the HLRS supercomputing center.
Many figures in this work were created using the VisIt code \citep{HPV:VisIt}.

\bibliographystyle{apj}
\bibliography{spirals}

\begin{thebibliography}{42}
\expandafter\ifx\csname natexlab\endcsname\relax\def\natexlab#1{#1}\fi

\bibitem[{{Beck} {et~al.}(1996){Beck}, {Brandenburg}, {Moss}, {Shukurov}, \&
  {Sokoloff}}]{Beck_1996}
{Beck}, R., {Brandenburg}, A., {Moss}, D., {Shukurov}, A., \& {Sokoloff}, D.
  1996, \araa, 34, 155

\bibitem[{{Bendre} {et~al.}(2015){Bendre}, {Gressel}, \&
  {Elstner}}]{Bendre_2015}
{Bendre}, A., {Gressel}, O., \& {Elstner}, D. 2015, Astronomische Nachrichten,
  336, 991

\bibitem[{{Bernet} {et~al.}(2008){Bernet}, {Miniati}, {Lilly}, {Kronberg}, \&
  {Dessauges-Zavadsky}}]{Bernet_2008}
{Bernet}, M.~L., {Miniati}, F., {Lilly}, S.~J., {Kronberg}, P.~P., \&
  {Dessauges-Zavadsky}, M. 2008, \nat, 454, 302

\bibitem[{{Brandenburg}(2015)}]{Brandenburg_2015}
{Brandenburg}, A. 2015, Astrophysics and Space Science Library, Vol. 407,
  {Simulations of Galactic Dynamos}, ed. A.~{Lazarian}, E.~M. {de Gouveia Dal
  Pino}, \& C.~{Melioli}, 529

\bibitem[{{Brandenburg} {et~al.}(2012){Brandenburg}, {Sokoloff}, \&
  {Subramanian}}]{Brandenburg_2012}
{Brandenburg}, A., {Sokoloff}, D., \& {Subramanian}, K. 2012, \ssr, 169, 123

\bibitem[{{Brandenburg} \& {Subramanian}(2005)}]{Brandenburg_2005}
{Brandenburg}, A., \& {Subramanian}, K. 2005, \physrep, 417, 1

\bibitem[{Childs {et~al.}(2012)Childs, Brugger, Whitlock, Meredith, Ahern,
  Pugmire, Biagas, Miller, Harrison, Weber, Krishnan, Fogal, Sanderson, Garth,
  Bethel, Camp, R\"{u}bel, Durant, Favre, \& Navr\'{a}til}]{HPV:VisIt}
Childs, H., Brugger, E., Whitlock, B., {et~al.} 2012, in {High Performance
  Visualization--Enabling Extreme-Scale Scientific Insight}, 357--372

\bibitem[{{Del Sordo} {et~al.}(2013){Del Sordo}, {Guerrero}, \&
  {Brandenburg}}]{DS_2013}
{Del Sordo}, F., {Guerrero}, G., \& {Brandenburg}, A. 2013, \mnras, 429, 1686

\bibitem[{{Donner} \& {Brandenburg}(1990)}]{Donner_1990}
{Donner}, K.~J., \& {Brandenburg}, A. 1990, \aap, 240, 289

\bibitem[{{Dubois} \& {Teyssier}(2008)}]{Dubois_Teyssier_2008}
{Dubois}, Y., \& {Teyssier}, R. 2008, \aap, 477, 79

\bibitem[{{Ferriere}(1998)}]{Ferriere_1998}
{Ferriere}, K. 1998, \aap, 335, 488

\bibitem[{{Fletcher} {et~al.}(2011){Fletcher}, {Beck}, {Shukurov},
  {Berkhuijsen}, \& {Horellou}}]{Fletcher_2011}
{Fletcher}, A., {Beck}, R., {Shukurov}, A., {Berkhuijsen}, E.~M., \&
  {Horellou}, C. 2011, \mnras, 412, 2396

\bibitem[{{Fromang} {et~al.}(2006){Fromang}, {Hennebelle}, \&
  {Teyssier}}]{Fromang_2006}
{Fromang}, S., {Hennebelle}, P., \& {Teyssier}, R. 2006, \aap, 457, 371

\bibitem[{{Gent} {et~al.}(2013){Gent}, {Shukurov}, {Sarson}, {Fletcher}, \&
  {Mantere}}]{Gent_2013}
{Gent}, F.~A., {Shukurov}, A., {Sarson}, G.~R., {Fletcher}, A., \& {Mantere},
  M.~J. 2013, \mnras, 430, L40

\bibitem[{{Gissinger} {et~al.}(2009){Gissinger}, {Fromang}, \&
  {Dormy}}]{Gissinger_2009}
{Gissinger}, C., {Fromang}, S., \& {Dormy}, E. 2009, \mnras, 394, L84

\bibitem[{{Gnedin} {et~al.}(2000){Gnedin}, {Ferrara}, \&
  {Zweibel}}]{Gnedin_2000}
{Gnedin}, N.~Y., {Ferrara}, A., \& {Zweibel}, E.~G. 2000, \apj, 539, 505

\bibitem[{{Gressel} {et~al.}(2008){Gressel}, {Elstner}, {Ziegler}, \&
  {R{\"u}diger}}]{Gressel_2008}
{Gressel}, O., {Elstner}, D., {Ziegler}, U., \& {R{\"u}diger}, G. 2008, \aap,
  486, L35

\bibitem[{{Hubbard} {et~al.}(2009){Hubbard}, {Del Sordo}, {K{\"a}pyl{\"a}}, \&
  {Brand enburg}}]{Hubbard_2009}
{Hubbard}, A., {Del Sordo}, F., {K{\"a}pyl{\"a}}, P.~J., \& {Brand enburg}, A.
  2009, \mnras, 398, 1891

\bibitem[{{Krause} \& {Raedler}(1980)}]{Krause_1980}
{Krause}, F., \& {Raedler}, K.~H. 1980, {Mean-field magnetohydrodynamics and
  dynamo theory}

\bibitem[{{Lopez-Rodriguez} {et~al.}(2020){Lopez-Rodriguez}, {Dowell}, {Jones},
  {Harper}, {Berthoud}, {Chuss}, {Dale}, {Guerra}, {Hamilton}, {Looney},
  {Michail}, {Nikutta}, {Novak}, {Santos}, {Sheth}, {Siah}, {Staguhn},
  {Stephens}, {Tassis}, {Trinh}, {Ward-Thompson}, {Werner}, {Wollack},
  {Zweibel}, \& {HAWC+Science Team}}]{Lopez-Rodriguez_2020}
{Lopez-Rodriguez}, E., {Dowell}, C.~D., {Jones}, T.~J., {et~al.} 2020, \apj,
  888, 66

\bibitem[{{Moffatt}(1978)}]{Moffatt_1978}
{Moffatt}, H.~K. 1978, {Magnetic field generation in electrically conducting
  fluids}

\bibitem[{{Navarro} {et~al.}(1996){Navarro}, {Frenk}, \& {White}}]{NFW96}
{Navarro}, J.~F., {Frenk}, C.~S., \& {White}, S.~D.~M. 1996, \apj, 462, 563

\bibitem[{{Ntormousi}(2018)}]{EN_2018}
{Ntormousi}, E. 2018, \aap, 619, L5

\bibitem[{{Pakmor} {et~al.}(2017){Pakmor}, {G{\'o}mez}, {Grand }, {Marinacci},
  {Simpson}, {Springel}, {Campbell}, {Frenk}, {Guillet}, {Pfrommer}, \&
  {White}}]{Pakmor_2017}
{Pakmor}, R., {G{\'o}mez}, F.~A., {Grand }, R. J.~J., {et~al.} 2017, \mnras,
  469, 3185

\bibitem[{{Parker}(1955)}]{Parker_1955}
{Parker}, E.~N. 1955, \apj, 122, 293

\bibitem[{{Parker}(1971)}]{Parker_1971}
---. 1971, \apj, 163, 255

\bibitem[{{Perret}(2016)}]{Perret_2016}
{Perret}, V. 2016, {DICE: Disk Initial Conditions Environment}, Astrophysics
  Source Code Library

\bibitem[{{Perret} {et~al.}(2014){Perret}, {Renaud}, {Epinat}, {Amram},
  {Bournaud}, {Contini}, {Teyssier}, \& {Lambert}}]{Perret_2014}
{Perret}, V., {Renaud}, F., {Epinat}, B., {et~al.} 2014, \aap, 562, A1

\bibitem[{{Rees}(1987)}]{Rees_1987}
{Rees}, M.~J. 1987, \qjras, 28, 197

\bibitem[{{Rheinhardt} \& {Brandenburg}(2010)}]{Rheinhardt_2010}
{Rheinhardt}, M., \& {Brandenburg}, A. 2010, \aap, 520, A28

\bibitem[{{Rieder} \& {Teyssier}(2016)}]{Rieder_Teyssier_2016}
{Rieder}, M., \& {Teyssier}, R. 2016, \mnras, 457, 1722

\bibitem[{{Rieder} \& {Teyssier}(2017)}]{Rieder_Teyssier_2017}
---. 2017, \mnras, 471, 2674

\bibitem[{{Shukurov} {et~al.}(2019){Shukurov}, {Rodrigues}, {Bushby},
  {Hollins}, \& {Rachen}}]{Shukurov_2019}
{Shukurov}, A., {Rodrigues}, L. F.~S., {Bushby}, P.~J., {Hollins}, J., \&
  {Rachen}, J.~P. 2019, \aap, 623, A113

\bibitem[{{Shukurov} {et~al.}(2006){Shukurov}, {Sokoloff}, {Subramanian}, \&
  {Brand enburg}}]{Shukurov_2006}
{Shukurov}, A., {Sokoloff}, D., {Subramanian}, K., \& {Brand enburg}, A. 2006,
  \aap, 448, L33

\bibitem[{{Steenbeck} \& {Krause}(1966)}]{Steenbeck_1966}
{Steenbeck}, M., \& {Krause}, F. 1966, Zeitschrift Naturforschung Teil A, 21,
  1285

\bibitem[{{Steenbeck} {et~al.}(1966){Steenbeck}, {Krause}, \&
  {R{\"a}dler}}]{SKR_1966}
{Steenbeck}, M., {Krause}, F., \& {R{\"a}dler}, K.~H. 1966, Zeitschrift
  Naturforschung Teil A, 21, 369

\bibitem[{{Steinwandel} {et~al.}(2019){Steinwandel}, {Beck}, {Arth}, {Dolag},
  {Moster}, \& {Nielaba}}]{Steinwandel_2019}
{Steinwandel}, U.~P., {Beck}, M.~C., {Arth}, A., {et~al.} 2019, \mnras, 483,
  1008

\bibitem[{{Stix}(1975)}]{Stix_1975}
{Stix}, M. 1975, \aap, 42, 85

\bibitem[{{Subramanian}(2016)}]{Subramanian_2016}
{Subramanian}, K. 2016, Reports on Progress in Physics, 79, 076901

\bibitem[{{Sutherland} \& {Dopita}(1993)}]{Sutherland_Dopita_1993}
{Sutherland}, R.~S., \& {Dopita}, M.~A. 1993, \apjs, 88, 253

\bibitem[{{Teyssier}(2002)}]{Teyssier_02}
{Teyssier}, R. 2002, \aap, 385, 337

\bibitem[{{Vainshtein} \& {Cattaneo}(1992)}]{Vainshtein_Cattaneo_1992}
{Vainshtein}, S.~I., \& {Cattaneo}, F. 1992, \apj, 393, 165

\end{thebibliography}

\end{document}